\documentclass[twocolumn,showpacs,preprintnumbers,amsmath,amssymb,prb,nofootinbib,superscriptaddress]{revtex4-2}

\usepackage{graphicx}
\usepackage{amsmath}
\usepackage{amssymb}
\usepackage{booktabs}
\usepackage{multirow}
\usepackage{xcolor}
\usepackage{hyperref}
\usepackage{bm}
\usepackage{physics}
\usepackage{microtype}

\newcommand{\obs}{x_\mathrm{obs}}
\newcommand{\param}{\theta}

\begin{document}

\title{Posterior Inference of Hamiltonian Parameters from RIXS Spectroscopy}

\author{Samuel Klein}
\affiliation{Machine Learning, SLAC National Accelerator Laboratory, Menlo Park, California 94025, USA}
\author{Thomas M. Linker}
\affiliation{Linac Coherent Light Source, SLAC National Accelerator Laboratory, Menlo Park, California 94025, USA}
\author{Louis Conreux}
\affiliation{Linac Coherent Light Source, SLAC National Accelerator Laboratory, Menlo Park, California 94025, USA}
\author{Daniel Ratner}
\affiliation{Computational Sciences and Technology Division, Thomas Jefferson National Accelerator Facility, Newport News, Virginia 23606, USA}
\affiliation{Department of Physics, Old Dominion University, Norfolk, Virginia 23529, USA}
\author{Apurva Mehta}
\affiliation{Linac Coherent Light Source, SLAC National Accelerator Laboratory, Menlo Park, California 94025, USA}
\author{Makoto Tachibana}
\affiliation{Research Center for Materials Nanoarchitectonics, National Institute for Materials Science, 1-1 Namiki, Tsukuba 305–0044, Ibaraki, Japan}
\author{Benjamin Zager}
\author{Jiemin Li}
\author{Jonathan~Pelliciari}
\author{Valentina~Bisogni}
\affiliation{National Synchrotron Light Source II, Brookhaven National Laboratory, Upton, New York 11973, USA}
\author{Wei He}
\affiliation{Stanford Institute for Materials and Energy Sciences,
SLAC National Accelerator Laboratory, Menlo Park, California 94025, USA}
\author{Xiangpeng Luo}
\affiliation{Condensed Matter Physics and Materials Science Department, Brookhaven National Laboratory, Upton, New York 11973, USA}
\author{Mark P. M. Dean}
\affiliation{Condensed Matter Physics and Materials Science Department, Brookhaven National Laboratory, Upton, New York 11973, USA}
\affiliation{Department of Physics and Astronomy, The University of Tennessee, Knoxville, Tennessee 37996, USA}
\author{Marton K.\ Lajer}
\affiliation{Condensed Matter Physics and Materials Science Department, Brookhaven National Laboratory, Upton, New York 11973, USA}

\author{Michael Kagan}
\affiliation{Fundamental Physics, SLAC National Accelerator Laboratory, Menlo Park, California 94025, USA}
\author{Joshua J. Turner}
\affiliation{Linac Coherent Light Source, SLAC National Accelerator Laboratory, Menlo Park, California 94025, USA}
\affiliation{Stanford Institute for Materials and Energy Sciences,
SLAC National Accelerator Laboratory, Menlo Park, California 94025, USA}
\author{Yongqiang Cheng}
\affiliation{Oak Ridge National Laboratory, Oak Ridge, Tennessee 37831, USA}
\author{Sean Gasiorowski}
\affiliation{Machine Learning, SLAC National Accelerator Laboratory, Menlo Park, California 94025, USA}

\date{March 2026} 

\begin{abstract}
We present the first application of simulation-based inference to resonant inelastic X-ray scattering spectroscopy. Using truncated marginal neural ratio estimation to efficiently restrict the prior and conditional flow matching as the joint density estimator, we infer full posteriors with a modest simulation budget for two Ni$^{2+}$ compounds---NiPS$_3$ as a representative covalent case and K$_2$NiF$_4$ as a more atomic one.
We demonstrate that a vision transformer encoder whose tokenization matches the physical layout of the RIXS map yields better-covered and sharper posteriors than generic image encoders.
Applying the validated method to experimental NiPS$_3$ and K$_2$NiF$_4$ data, we recover a joint posterior that reveals parameter correlations invisible to point estimators, and a posterior predictive distribution that closely matches the observed spectrum.
The amortized posterior unlocks a class of analyses not previously available to the field such as nuisance-marginalized uncertainty quantification, multi-measurement posterior fusion and active experimental design.
\end{abstract}

\maketitle

\section{Introduction}

Resonant inelastic X-ray scattering (RIXS) has emerged as a powerful probe of quantum materials, providing deep insight into correlated electron physics. 
RIXS is a photon-in photon-out process where an incident X-ray promotes a core electron into a resonant excited state, which then subsequently decays and emits a second photon. 
The energy and momentum transfer of this decay process is imprinted on the second photon and provides direct access to charge, spin, and orbital excitations of the material with elemental specificity as well as low energy collective excitations such as phonons, magnons, and excitons~\cite{Ament2011,Kotani2001,Dean2015,deGroot2024,Mitrano2024exploring},    

For most quantum materials, interpretation of RIXS spectra is done through semi-empirical Hamiltonians, as the locality and many-body nature of X-ray excitations are poorly described by most \textit{ab-initio} methods for extended systems~\cite{Kotani2001,deGroot2008}.
Extracting the empirical parameters for these Hamiltonians can provide deep insight into their overall properties and function. 
For example, transition metal RIXS spectra at the $L_{2,3}$-edge(2p$\xrightarrow{}$ 3d,4d,5d) is strongly imprinted by $dd$ excitations which reflect the multi-orbital Hamiltonian crystal field splitting, Coulomb interactions, and spin-orbit coupling among the $d$ electrons~\cite{deGroot2008}. 
Quantitative understanding of the Hamiltonian parameters that reproduce such excitations constrains models for magnetic ordering, exciton formation, and charge-transfer processes in these materials~\cite{Haverkort2012}.

Extracting the parameters of semi-empirical Hamiltonians from RIXS spectra is a challenging inverse problem. 
The forward map from parameters to spectrum is nonlinear, expensive to evaluate, requires codes such as EDRIXS~\cite{EDRIXS2019}, and multiple parameter sets can yield similar spectra.
Combined with measurement noise, this means the parameters cannot be determined uniquely, and the scientifically meaningful output is a joint probability distribution over parameters rather than a single best fit.

Historically, parameter extraction has relied on expert manual tuning, a labor-intensive process that provides no formal uncertainty quantification and is difficult to reproduce. Recent work by \textcite{Lajer2025} has automated this process using Bayesian optimization, where a surrogate for the $L_1$ spectral distance is iteratively refined, guiding the simulation budget toward promising parameter regions, with results subsequently refined via local gradient-free optimization. 
The method is a significant step forward, demonstrating systematic, reproducible parameter extraction.
The authors report best-fit parameters and empirical confidence intervals defined as the parameter range within a fixed multiplicative factor of the minimum distance. 
These intervals, however, are univariate, and do not capture correlations between parameters. Further, the procedure relies on a sensible, but ad hoc choice of an $L_1$ distance as an optimization target. Minimizing an $L_1$ or $L_2$ norm corresponds to maximum-likelihood estimation under a Laplace or Gaussian likelihood, respectively.

We adopt simulation-based inference (SBI), which replaces this implicit likelihood with an explicit statistical model that places a posterior distribution $p(\param \mid \obs)$ over Hamiltonian parameters $\param$ given the observed spectrum $\obs$.
To apply SBI to the RIXS problem, we focus on simulation efficiency through truncated marginal neural ratio estimation (TMNRE)~\cite{miller2021truncated}, and introduce a novel column vision transformer (column-ViT) embedding network whose patch structure matches the physical layout of the RIXS measurement, enabling data-efficient training in the small-simulation-budget regime that expensive EDRIXS evaluations impose.

The approach developed in this work provides several advantages over $L_1$ optimization: (i)~a joint distribution over all parameters of interest that captures correlations between parameters; (ii)~uncertainty quantification that marginalizes over nuisance parameters, such as instrumental parameters; and (iii)~amortization, which means that once the initial training cost is paid, the learned posterior can be applied almost instantly to additional measurements of materials sharing the same or similar prior structure.

The paper is organized as follows. Section~\ref{sec:methods} introduces the RIXS forward model, the SBI framework, the TMNRE pipeline, the column-ViT encoder, and the nuisance-marginalization scheme. Section~\ref{sec:results} applies the method to two materials spanning a range of bonding character, NiPS$_3$, a more covalent quantum material for which we compare against the \textcite{Lajer2025} reference simulation, and K$_2$NiF$_4$, a more atomic case. Section~\ref{sec:conclusion} concludes and discusses limitations of the approach.

\section{Methods}
\label{sec:methods}

\subsection{RIXS Forward Model}

The simulation we use follows the single-ion model of \textcite{Lajer2025}, where the system is modeled as an isolated transition metal atom embedded in a crystal field of the lattice. The Hamiltonian for the ground state ($H_{i}$) and intermediate excited state ($H_{n}$) can be written as:
\begin{equation*}
\hat{H}
=
\hat{H}_{\mathrm{cf}}
+
\hat{H}_{\mathrm{soc}}
+
\hat{H}_{\mathrm{Coul}},
\end{equation*}

with 

\begin{align*}
\hat{H}_{\mathrm{cf}}
&=
\sum_{\alpha\beta}
\Delta_{\alpha\beta}
\hat{v}^{\dagger}_{\alpha}
\hat{v}^{\phantom{\dagger}}_{\beta},
\\
\hat{H}_{\mathrm{soc}}
&=
\zeta_{v}
\sum_{\alpha\beta}
\langle \alpha | \mathbf{l}\cdot\mathbf{s} | \beta \rangle
\hat{v}^{\dagger}_{\alpha}
\hat{v}^{\phantom{\dagger}}_{\beta}
+
\zeta_{c}
\sum_{\mu\nu}
\langle \mu | \mathbf{l}\cdot\mathbf{s} | \nu \rangle
\hat{c}^{\dagger}_{\mu}
\hat{c}^{\phantom{\dagger}}_{\nu},
\\
\hat{H}_{\mathrm{Coul}}
&=
\sum_{\alpha\beta\gamma\delta}
U^{vv}_{\alpha\beta\gamma\delta}
\hat{v}^{\dagger}_{\alpha}
\hat{v}^{\dagger}_{\beta}
\hat{v}^{\phantom{\dagger}}_{\delta}
\hat{v}^{\phantom{\dagger}}_{\gamma}
+
\sum_{\alpha\mu\beta\nu}
U^{vc}_{\alpha\mu\beta\nu}
\hat{v}^{\dagger}_{\alpha}
\hat{c}^{\dagger}_{\mu}
\hat{c}^{\phantom{\dagger}}_{\nu}
\hat{v}^{\phantom{\dagger}}_{\beta}.
\end{align*}

where $\hat{v}^{\dagger}_{\alpha}$ and $\hat{c}^{\dagger}_{\alpha}$ represent creation operators for the valence ($d$ orbitals) and core orbitals ($p$ orbitals) respectively, and where $\alpha$ labels specific spin-orbitals. 
The crystal field splitting is then encoded in $\Delta_{\alpha\beta}$, spin-orbit coupling in $\zeta_{v}$ and $\zeta_{c}$, and Coulomb interactions in $U^{vc}_{\alpha\mu\beta\nu}$ and $U^{vv}_{\alpha\beta\gamma\delta}$. 
The crystal field interaction is often encoded in terms of a single value 10$Dq$ and the symmetry of the system, while the Coloumb interactions are written in terms of Slater integrals: $F^{(2)}$,$F^{(4)}$,$G^{(1)}$,$G^{(3)}$. For simplicity, when computing $H_{i}$ terms involving the core orbitals are neglected as they only result in an additive constant. The input Hamiltonian is parameterized by the quantities listed in Table~\ref{tab:params}.

 The RIXS cross section $I(\omega_{\mathrm{in}},\Omega \equiv \omega_{\mathrm{in}} - \omega_{\mathrm{out}})$ is then computed via the Kramers-Heisenberg equation~\cite{Ament2011,deGroot2024} using exact diagonalization of the valence and core electron states:
\begin{align*}
&I(\omega_{\mathrm{in}},\Omega)
=
\sum_{i,f}
\frac{e^{-E_i/k_B T}}{Z}
\left| M_{fi}(\omega_{\mathrm{in}}) \right|^2
\delta\!\left(E_f - E_i - \hbar\Omega\right),
\\[6pt]
&M_{fi}(\omega_{\mathrm{in}})
=
\sum_n
\frac{
\langle f | \hat{D}_{\mathrm{out}}^{\dagger} | n \rangle
\langle n | \hat{D}_{\mathrm{in}} | i \rangle
}{
E_i + \hbar\omega_{\mathrm{in}} - E_n + i\Gamma_n/2
},
\\[6pt]
&\hat{H}_i | i \rangle = E_i | i \rangle,
\qquad
\hat{H}_n | n \rangle = E_n | n \rangle,
\qquad
\hat{H}_f | f \rangle = E_f | f \rangle.
\end{align*}
 For the RIXS processes investigated here, the final state and ground state Hamiltonians are the same.  The forward model is implemented in EDRIXS~\cite{EDRIXS2019}, an open-source package for core-hole spectroscopy simulations. Following \textcite{Lajer2025}, all simulated and observed spectra are sum-normalized before analysis, removing overall intensity as a degree of freedom.

\begin{table}[t]
\caption{Hamiltonian parameters used in the EDRIXS forward model. Unless otherwise noted parameters are kept the same in the initial and final states}
\label{tab:params}
\begin{ruledtabular}
\begin{tabular}{ll}
Parameter & Physical meaning \\
\hline
$F^{(2)}_{dd}$, $F^{(4)}_{dd}$ & $d$--$d$ Coulomb Slater integrals \\
$F^{(2)}_{dp}$, $G^{(1)}_{dp}$, $G^{(3)}_{dp}$ & $d$--$p$ Slater integrals \\
$\zeta_{v,i}$, $\zeta_{v,n}$ & Valence spin-orbit coupling\\
$\zeta_c$ & Core spin-orbit coupling \\
$10Dq$ & Crystal field splitting \\
$\Gamma_c$ & Core-hole lifetime broadening \\
$\Delta\omega_{\mathrm{in}}$ & Incident energy offset \\
$\sigma$ & Energy-loss broadening \\
\end{tabular}
\end{ruledtabular}
\end{table}

\subsection{Simulation-Based Inference}

SBI~\cite{Cranmer2020} is a family of methods for Bayesian inference targeting applications where the likelihood $p(x \mid \param)$ is intractable but the forward model can generate simulated data $x$ for any parameter vector $\param$, producing pairs $(\param, x)$. 
In Bayesian inference the goal is to compute the posterior distribution $p(\param \mid \obs) \propto p(\obs \mid \param) p(\param)$, where $p(\param)$ is a prior over parameters and $p(\obs \mid \param)$ is the likelihood of the observed data given the parameters.
Our approach proceeds in two stages.
In the first stage, truncated marginal neural ratio estimation (TMNRE)~\cite{miller2021truncated} efficiently identifies the region of parameter space where the posterior has appreciable support, starting from a broad prior and iteratively concentrating the simulation budget toward high-posterior regions.
In the second stage, a joint density estimator is trained on simulations drawn from this restricted region, directly approximating the full multivariate posterior.
This two-stage design separates the problem of finding where the posterior lives from the problem of accurately modelling its shape, allowing each stage to use the most appropriate tool.

In neural ratio estimation (NRE)~\cite{Hermans2020}, a binary classifier is trained to distinguish joint pairs $(\param, x) \sim p(\param, x)$ from marginal pairs $(\param, x) \sim p(\param)\,p(x)$.
The optimal classifier output is related to the likelihood-to-evidence ratio
\begin{equation*}
r(\param, x) = \frac{p(x \mid \param)}{p(x)},
\end{equation*}
where $p(x \mid \param)$ is the likelihood of observing data $x$ given parameters $\param$ and $p(x) = \int p(x \mid \param)\,p(\param)\,d\param$ is the evidence, the marginal probability of the data averaged over the prior.
This ratio quantifies how much more likely the observed data are under a specific parameter value than under the prior as a whole. TMNRE~\cite{miller2021truncated} trains such classifiers independently for each marginal $p(\theta_j \mid x)$ and uses the learned ratios to iteratively truncate the prior, discarding regions where the marginal ratio falls below a threshold and drawing new simulations only from the restricted support.
Concretely, new simulations are generated by rejection sampling from the prior within the tightened bounds for each parameter.
This concentrates the simulation budget near the high-posterior region, dramatically improving efficiency relative to sampling uniformly from the full prior.
Because each marginal classifier operates on a single scalar parameter, the method scales favourably with parameter count and remains effective even from a broad, uninformative prior.

Once TMNRE has identified the high-posterior region, a joint density estimator $q_\phi(\param \mid x)$ is trained on simulations drawn from the restricted prior to directly approximate the posterior---an approach known as neural posterior estimation~\cite{Papamakarios2016}. In this work the density estimator is a conditional flow matching model~\cite{Lipman2023,Albergo2023} that learns a time-dependent velocity field transporting samples from a Gaussian base distribution to the posterior via an ordinary differential equation. The conditional structure allows the same network to produce posterior samples for any observed spectrum after training, enabling amortized inference across multiple experimental datasets without retraining.

A secondary benefit of the SBI framework is nuisance marginalization at no additional implementation cost. We train the density estimator to learn the conditional posterior $q_\phi(\param \mid \obs, \sigma)$, treating the energy-loss broadening $\sigma$ as a conditioning input. At inference time we compute the marginalized posterior $p(\param \mid \obs) = \int q_\phi(\param \mid \obs, \sigma)\,p(\sigma)\,d\sigma$, integrating over the prior on $\sigma$.
In practice, sampling from this marginalized posterior requires only drawing $\sigma \sim p(\sigma)$ and then drawing $\param \sim q_\phi(\param \mid \obs, \sigma)$, so the marginalization adds no computational overhead beyond what is already required to produce posterior samples. This construction encodes the assumption that the spectra contain no information about the energy-loss broadening beyond external calibration, that is, $p(\sigma \mid \obs) = p(\sigma)$.
We discuss the justification and potential failure modes of this assumption in Appendix~\ref{app:sigma-assumption}. 
Standard $L_1$ optimization fixes $\sigma$ to a single point estimate. This can bias parameters whose posteriors shift with the assumed broadening, and it produces uncertainties conditional on an unvalidatable choice. Marginalizing over $p(\sigma)$ removes both issues at no additional cost once the density estimator conditions on $\sigma$. For well-calibrated instruments the quantitative effect is modest (Appendix~\ref{app:sigma-assumption}), but the marginalization guarantees that reported uncertainties reflect calibration uncertainty rather than ignoring it.

\begin{figure*}[t]
    \centering
    \includegraphics[width=\linewidth]{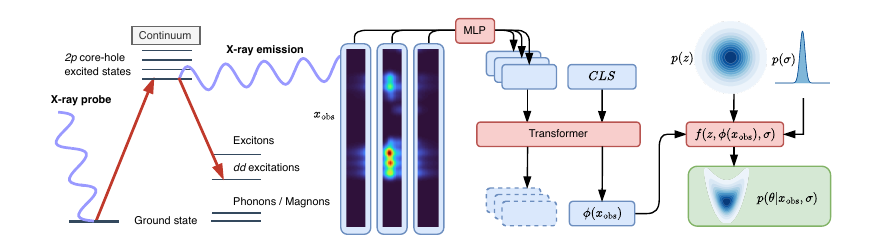}
    \caption{Overview of the experimental and inference pipeline. An incident X-ray photon resonantly excites a $2p$ core electron into an unoccupied valence state, creating a core-hole intermediate state that decays as a valence electron refills the hole and a scattered photon is emitted, leaving the system in a $dd$-excited final state. The energy loss of the scattered photon encodes the excitation energy, and scanning over incident energies builds up the observed 2D spectrum $\obs$. The column-ViT encoder divides $\obs$ into column-wise patches along the incident-energy axis, embeds each patch with a shared Multi-Layer Perceptron (MLP), and passes the resulting tokens together with a learnable $\mathit{CLS}$ token through a transformer to produce a context vector $\phi(\obs)$. A conditional flow matching model, conditioned on the context vector $\phi(\obs)$ and $\sigma\sim p(\sigma)$, transports samples from a Gaussian base density into the posterior $p(\param \mid \obs)$ by integrating a learned velocity field. The column-ViT encoder and velocity network $f$ are trained jointly on simulated parameter-spectrum pairs.}
    \label{fig:pipeline}
\end{figure*}

\subsection{Prior Distribution}
\label{sec:prior}

In SBI the prior $p(\boldsymbol{\theta})$ encodes what is known about the parameters before the data are observed and determines the region of parameter space over which the posterior has support. Choosing $p(\boldsymbol{\theta})$ is the natural place to inject physical knowledge into the inference. For isolated atoms of a given $d$ electron configuration the values of the parameters describing the Coulomb interaction and spin-orbit coupling are exactly known. Deviations from these values reflect the electronic interaction and screening of the atom within the lattice.  We  exploit this fact by anchoring the bounds for most parameters based on their tabulated atomic values, which provide a well-defined reference point from which deviations due to lattice interactions can be motivated. For all 12 parameters we place uniform priors over these physically motivated bounds, softened with a Tukey window so that the flat central region tapers smoothly to zero at each boundary rather than cutting off sharply.

\paragraph{Slater integrals.}
The five Slater integrals ($F^{(2)}_{dd}$, $F^{(2)}_{dp}$, $F^{(4)}_{dd}$, $G^{(1)}_{dp}$, $G^{(3)}_{dp}$) are bounded relative to their atomic values.
In bound states, electron--electron repulsion integrals are systematically reduced from their atomic values due to orbital-expansion effects.
The reduction is typically 20--40\% for transition-metal compounds~\cite{Cowan1981,deGroot2008}.
We set Tukey-windowed uniform bounds of $[0.3\,P_\text{at},\;1.1\,P_\text{at}]$ for $F^{(2)}_{dd}$, $F^{(2)}_{dp}$, $G^{(1)}_{dp}$, and $G^{(3)}_{dp}$, allowing a generous reduction below the atomic value while permitting modest upward variation. Rather than drawing $F^{(4)}_{dd}$ independently, we sample a ratio $r \sim \mathcal{N}(0.625,\,0.1)$ and set $F^{(4)}_{dd} = r \cdot F^{(2)}_{dd}$, concentrating the joint prior along the atomic ratio of $F^{(4)}_{dd}/F^{(2)}_{dd}=0.625$ observed across the $3d$ series.

\paragraph{Spin--orbit coupling.}
The three spin-orbit coupling constants $\zeta_c$, $\zeta_{v,i}$, $\zeta_{v,n}$ are given Tukey-windowed uniform priors over $[0.8\,\zeta_\text{at},\;1.5\,\zeta_\text{at}]$.
The lower bound allows for moderate reduction due to covalency effects, while the upper bound accommodates the possibility that effective spin-orbit coupling in the solid exceeds the free-atom value.

\paragraph{Core-hole broadening and energy-loss broadening.}
The core-hole lifetime broadening $\Gamma_c$ and the energy-loss broadening $\sigma$ are the two parameters whose bounds are set by the individual measurement rather than by atomic reference values, and they differ between the two materials. The parameter $\sigma$ absorbs both the instrumental energy resolution and the finite lifetime of the $dd$ excitations, which together govern the observed peak widths along the energy-loss axis. For NiPS$_3$ we place a Tukey-windowed uniform prior on $\Gamma_c$ centered on $0.57\,\text{eV}$ with a half-width of $0.2\,\text{eV}$, consistent with experimentally reported Ni $2p$ values, and take $\sigma \in [0.030,\;0.036]\,\text{eV}$, reflecting the combination of the reported instrumental resolution and expected excitation lifetimes, with bounds reflecting our estimated calibration uncertainty. For K$_2$NiF$_4$ we use the broader ranges $\Gamma_c \in [0.1,\;0.8]\,\text{eV}$ and $\sigma \in [0.01,\;0.05]\,\text{eV}$, reflecting the weaker external constraint available for that measurement.
The energy-loss broadening $\sigma$ is treated as an externally calibrated nuisance parameter. Because instrumental resolution and intrinsic excitation-lifetime broadening produce similar spectral signatures, they cannot be reliably disentangled from a single RIXS spectrum without independent calibration data. We therefore assume that the RIXS spectrum provides no additional constraint on $\sigma$ beyond the beamline calibration.

\paragraph{Crystal-field splitting and energy offset.}
The crystal-field parameter $10Dq$ is given a Tukey-windowed uniform prior over $[0.1,\;5.0]\,\text{eV}$, spanning the range observed across Ni oxides, halides, and chalcogenides.~\cite{shannon1976revised,van1988multiplet}.
The incident energy offset $\Delta\omega_{\mathrm{in}}$ is given a Tukey-windowed uniform prior over $[-10,\;10]\,\text{eV}$ to account for absolute-energy calibration uncertainties.

\subsection{Density estimator architecture}
\label{sec:encoder}

The flow matching model generates samples of the Hamiltonian parameters $\param$, conditioned on the RIXS map and the energy-loss broadening $\sigma$.
To condition on the spectrum, the 2D map is compressed into a vector representation by an encoder network, which is trained jointly with the velocity network.
The full pipeline from photons exciting the sample to the posterior distribution over parameters is shown in Fig.~\ref{fig:pipeline}.
There is considerable freedom in choosing the encoder in this process. Standard high-capacity image encoders such as ResNets~\cite{He2016} or Vision Transformers with patch-based tokenization~\cite{Dosovitskiy2021} treat the input as a generic 2D image and learn all relevant structure from data. Although patch-based Vision Transformers include positional encodings that break strict translation equivariance, their square-patch tokenization does not reflect the distinct physical roles of the two spectral axes, and the positional information they encode is learned from data rather than built into the tokenization structure. We instead prefer an architecture whose inductive bias mirrors the physical structure of the RIXS map, both because physically grounded structure tends to generalize better and because the simulation budget is below what a generic high-capacity encoder would need.

We retain the Vision Transformer framework~\cite{Vaswani2017,Dosovitskiy2021} but replace the standard square-patch tokenization with one that respects the distinct physical roles of the two spectral axes. The central design choice is then how the 2D map is divided into discrete input tokens before attention-based aggregation. In this work column tokenization refers to assigning one token per incident-energy slice, row tokenization one per energy-loss slice, and flat tokenization treats the entire map as a single vector.
The two tokenized schemes retain absolute position through their positional encodings, so the choice between column and row tokenization is the axis compressed to form the tokens.
The resulting column Vision Transformer (column-ViT) divides the spectrum into column-wise patches along the incident-energy axis, embeds each patch with a shared Multi-Layer Perceptron (MLP), and passes the resulting tokens together with a learnable classification ($\mathit{CLS}$) token~\cite{Devlin2019} through a transformer encoder; the transformed $\mathit{CLS}$ token is read out as a 128-dimensional context vector $\phi(\obs)$. The transformer uses 4 layers of 4-head attention with GELU activations, an MLP expansion ratio of 4, dropout of 0.2 during training, and 1D sinusoidal positional embeddings along the incident-energy axis.
The energy-loss broadening $\sigma$ is also a conditioning variable of the density estimator. 
It enters the network by normalizing to $[-1, 1]$ over the training-set range and concatenating the resulting scalar to the encoder output, so the full context passed to the velocity network is $[\phi(\obs),\,\tilde\sigma]$.
Each column corresponds to a fixed incident energy and independently probes a subset of resonant transitions, so the inductive bias of patch-wise attention along this axis is physically motivated.
The advantage over other tokenizations and aggregation strategies is demonstrated in Section~\ref{sec:encoder-selection}.

For the density estimator we use conditional flow matching with an optimal-transport interpolant~\cite{Lipman2023,Albergo2023}. The velocity network is a four-layer MLP with 256 hidden units per layer and SiLU activations, taking as input the concatenation of the current state $\param_t$, the time $t$, and the encoder context vector $\phi(\obs)$, and outputting a velocity of the same dimension as $\param$. Training minimizes the mean-squared error between the predicted velocity and the optimal-transport target $\param - z$, where $z \sim \mathcal{N}(0, I)$ is the noise sample. At inference time, posterior samples are obtained by integrating the learned velocity field from $t=0$ to $t=1$ with 100 Euler steps starting from Gaussian noise. An exponential moving average of the velocity network weights over training iterations is used for validation and final sampling. Further training details are given in Appendix~\ref{app:flow-matching-details}.
We train a deep ensemble~\cite{Lakshminarayanan2017} of independently initialized flow matching models on the same dataset and pool their posterior samples at inference time, guarding against overconfident posteriors from any single approximate model~\cite{Hermans2022crisis}.

\subsection{Active Learning and Sequential Refinement}

In the TMNRE stage, the prior is iteratively restricted by training one-dimensional binary classifiers for each parameter independently. As each classifier operates on a single scalar parameter value concatenated with a low-dimensional summary of the spectrum, this stage requires few simulations per round and scales favorably with the number of parameters. We train an ensemble of five classifiers per parameter, all instantiated as a single batched model whose weights are updated simultaneously via batched matrix multiplications, making the per-round training cost negligible compared to the simulation cost.

For training on small datasets we use a low-dimensional summary of the 2D RIXS map rather than the full spectrum. 
This is done by applying PCA to the flattened spectra, retaining components up to 95\% cumulative explained variance. These PCA scores are concatenated with the first and second moments of the energy-loss profile computed in 10 slices along the incident-energy axis, providing complementary information about peak positions and their broadening. Full details of the summary construction are given in Appendix~\ref{app:summaries}.

Each round, parameters whose marginal log-ratio falls below $\epsilon = 10^{-6}$ of the peak are truncated from the prior, and new simulations are drawn uniformly from the restricted bounds. We run eight rounds of truncation with 1{,}000 simulations per round, each round training on all data accumulated up to that point. The truncation threshold corresponds to approximately $\pm 5.26\sigma$ for a Gaussian marginal, ensuring that the restricted prior remains conservative. At each round we check that highest-posterior-density intervals built from the marginal ratio estimators remain close to their nominal coverage, confirming that the threshold is not discarding regions the data support (Appendix~\ref{app:sbc}).

Once the prior has been restricted, 100{,}000 simulations are drawn from the final restricted prior and used to train the joint density estimator. The encoder and velocity network are trained jointly with early stopping on a held-out validation set. Full training hyperparameters are given in Appendix~\ref{app:flow-matching-details}. The total simulation budget of 108{,}000 is the same order of magnitude as the approximately 60{,}000 evaluations used by \textcite{Lajer2025} for Bayesian optimization, though SBI additionally requires training the flow matching model and encoder.

The resulting restricted-prior dataset is agnostic to the choice of joint density estimator, so any conditional generative model can be trained on the same simulations without repeating the active learning stage.

\subsection{Posterior Analysis}

We run the eigenstate annotation code of \textcite{Lajer2025} to label the ground state and lowest excitations by irreducible representation of the local symmetry group at the posterior mean $\bar{\param} = \mathbb{E}[\param \mid \obs]$ and at a random draw of posterior samples, which propagates parameter uncertainty to derived physical labels.
We compare the observed data against the forward simulation at the parameter-space posterior mean $\bar{\param}$, and the posterior predictive mean spectrum $\bar{x} = \mathbb{E}[x^* \mid \obs]$ obtained by averaging forward simulations across posterior draws. The former provides an interpretable single parameter vector while the latter accounts for the nonlinearity of the forward model by averaging over posterior uncertainty.
We also study the posterior predictive distribution $p(x^* \mid \obs) = \int p(x^* \mid \param)\,p(\param \mid \obs)\,d\param$ as it gives the distribution of new observations implied by the posterior, marginalizing over parameter uncertainty. 
We approximate the posterior predictive by drawing parameter sets $\param^{(i)} \sim p(\param \mid \obs)$ and forward-simulating each through EDRIXS to produce an ensemble of predicted spectra. Comparing the resulting credible band against the measured spectrum tests whether the full posterior is consistent with the data. 
A posterior that is broad in individual parameters yet produces a tight predictive band signals spectral degeneracy.

\section{Results}
\label{sec:results}

\subsection{NiPS\texorpdfstring{$_3$}{3} data}
\label{sec:nips3}

The first test case of SBI models for Hamiltonian inference is the $L$-edge RIXS spectrum of NiPS$_3$, a layered van der Waals antiferromagnet hosting Ni$^{2+}$ ($d^8$) in a nearly octahedral environment~\cite{Wildes2015,Kang2020}.
The experimental spectrum was measured by \textcite{He2024} at the SIX 2-ID beamline of the National Synchrotron Light Source II (NSLS-II) with an energy resolution of 31~meV full width at half maximum. The spectrum was recorded at $T=40$~K with an incident x-ray angle $\theta_{\mathrm{in}}=23^{\circ}$, a scattering angle $2\Theta=150^{\circ}$, and $\pi$-polarized incident x-rays over an incident-energy range spanning the Ni $L_3$-edge. The measured spectrum was truncated to the energy-loss window $E_{\mathrm{loss}}\in[0.5,\,2.0]$~eV and resampled onto an equidistant grid of $40\times151$ points in incident energy and energy loss, respectively.
Its RIXS spectrum features prominent $dd$ excitations and a Hund's exciton near 1.45~eV that is not fully captured within the single-ion model~\cite{Kang2020,Lajer2025,He2024}.
We compare against two reference solutions. 
The first is a `hand-fit' parameter set reported by \textcite{He2024}. 
The second is from the automated analysis of \textcite{Lajer2025}, which combines Bayesian optimization of a Gaussian process surrogate with derivative-free local refinement~\cite{cartis2018improvingflexibilityrobustnessmodelbased} to minimize the $L_1$ spectral distance.

\subsubsection{Encoder Architecture Selection}
\label{sec:encoder-selection}

Every candidate encoder is trained on the same 100{,}000 restricted-prior simulations used for the results reported below. Each trained model is assessed using two criteria on 5{,}000 held-out simulations drawn from the restricted prior. The first criterion is whether the joint posterior has correct coverage, which we quantify with tests of accuracy with random points (TARP)~\cite{Lemos2023}. TARP measures how often the true parameter vector lies closer to a randomly drawn reference point than the posterior samples do, and we summarize it as the deviation of the resulting credibility distribution from uniformity, so that a posterior with perfect coverage scores zero. The second is sharpness, defined as the width of the 90\% credible interval averaged over parameters and normalized by the width of the restricted prior. Neither criterion selects an architecture on its own, since a model that returns the prior unchanged has perfect coverage while carrying no information, and a model can be arbitrarily sharp and biased. We therefore require a candidate to have correct coverage and rank the covered candidates by sharpness.

At matched embedding dimension, depth, dropout, and token count, the column-ViT achieves a TARP deviation of 0.060 and a sharpness of 0.22, while both the row-ViT and the flat MLP baseline reach a TARP deviation of 0.077 with sharpness of 0.38 and 0.34 respectively. Column tokenization therefore has both the best coverage and the sharpest posteriors. With the token budget fixed, the tokenization determines how strongly each axis is compressed before attention is applied, and the RIXS map is sampled far more finely in energy loss than in incident energy. Column tokenization therefore collapses only a few adjacent incident energies into each token while carrying the full energy-loss profile within it, whereas row tokenization must compress many energy-loss bins into every token. Each column token then spans a narrow range of incident energy and holds a complete spectrum for it, which is also the unit in which the measurement is assembled.
The full grid, the comparison between deterministic and stochastic inference, and the ensemble comparison are given in Appendix~\ref{app:encoder-ablation}.


\subsubsection{Posterior predictive}
\label{sec:ppc-validation}
 
The posterior mean Hamiltonian parameters capture the main features of the data, following the same overall shape and correctly reproducing the energy-loss positions of the $dd$ excitations as shown in Fig.~\ref{fig:spectrum_panel}. The numerical values of the posterior mean are given in Appendix~\ref{app:posterior-mean}.
Forward simulation of the \textcite{He2024} and \textcite{Lajer2025} reference parameters produce visually similar spectra to the posterior mean reconstruction, and all three are similar to the observed data.

\begin{figure}[htbp]
    \centering
    \includegraphics[width=1\linewidth]{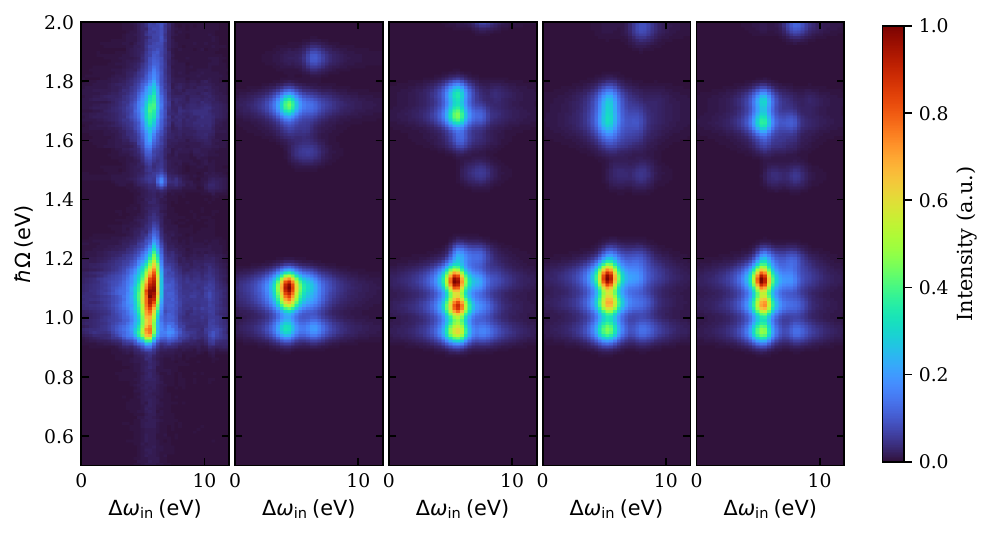}
    \caption{\emph{Far left:} the experimentally measured NiPS$_3$ RIXS spectrum. \emph{Center left:} the simulated spectrum at the by-hand reference values. \emph{Center:} the simulated spectrum at the \textcite{Lajer2025} reference parameter values after fine-tuning. \emph{Center right:} the posterior predictive spectral mean. \emph{Far right:} the simulated spectrum evaluated at the posterior mean of the SBI posterior.}
    \label{fig:spectrum_panel}
\end{figure}

\begin{figure*}[htbp]
\centering 
\includegraphics[width=1.0\linewidth]{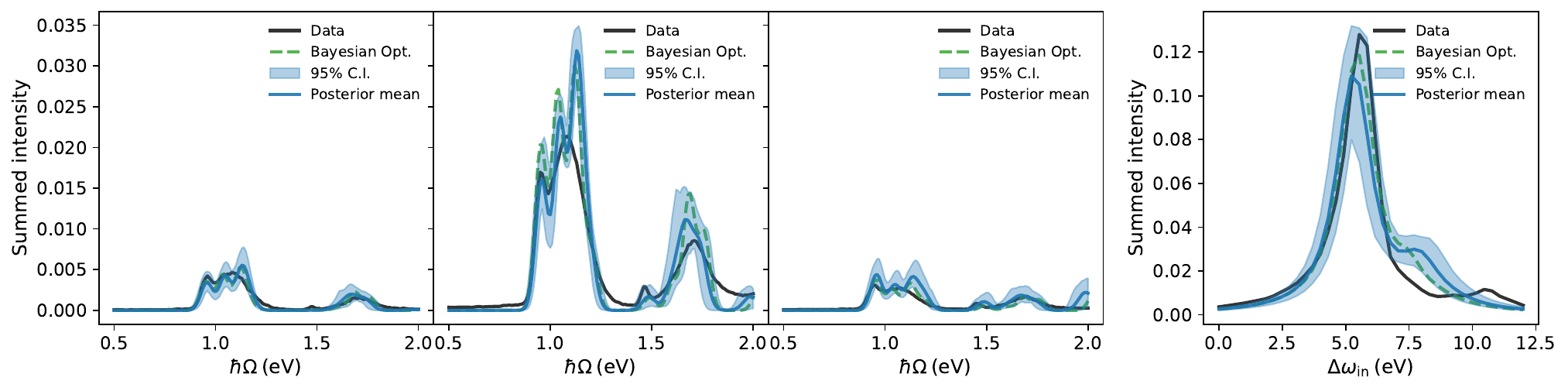}
\caption{Posterior predictive check for NiPS$_3$. Parameter sets drawn from the posterior are forward-simulated with EDRIXS and projected onto one-dimensional slices. Left panels show three non-overlapping $\Delta\omega_{\mathrm{in}}$ bins of width 4\,eV ($[0,4]$, $[4,8]$, and $[8,12]$\,eV, left to right); the right panel shows summed intensity over all $\hbar\Omega$ as a function of $\Delta\omega_{\mathrm{in}}$. Shaded bands span the 95\% credible interval of the predictive distribution, the solid black curve is the observed data, and the dashed green line is the Bayesian optimization result~\cite{Lajer2025}.}
\label{fig:ppc}
\end{figure*}

To visualize the full posterior predictive we project 2D spectra onto one-dimensional slices along the incident-energy axis. 
Figure~\ref{fig:ppc} shows the observed spectrum overlaid on the posterior predictive band formed from the simulated draws. 
The simulated spectra are concentrated, and approximate the mode locations and their relative intensities across different slices well in all slices, as does the result of \textcite{Lajer2025}.
The forward simulation at the posterior mean achieves an $L_1$ distance of $0.476$, modestly smaller than the $0.480$ reported by \textcite{Lajer2025} after local refinement, while the posterior predictive mean spectrum has an $L_1$ distance of $0.484$. The best individual posterior sample achieves $0.472$, demonstrating that the posterior concentrates around spectra at least as close to the data as the optimized point estimate.

\subsubsection{Marginal posterior}
\label{sec:marginals}

\begin{figure*}[htbp]
    \centering
    \includegraphics[width=1.0\linewidth]{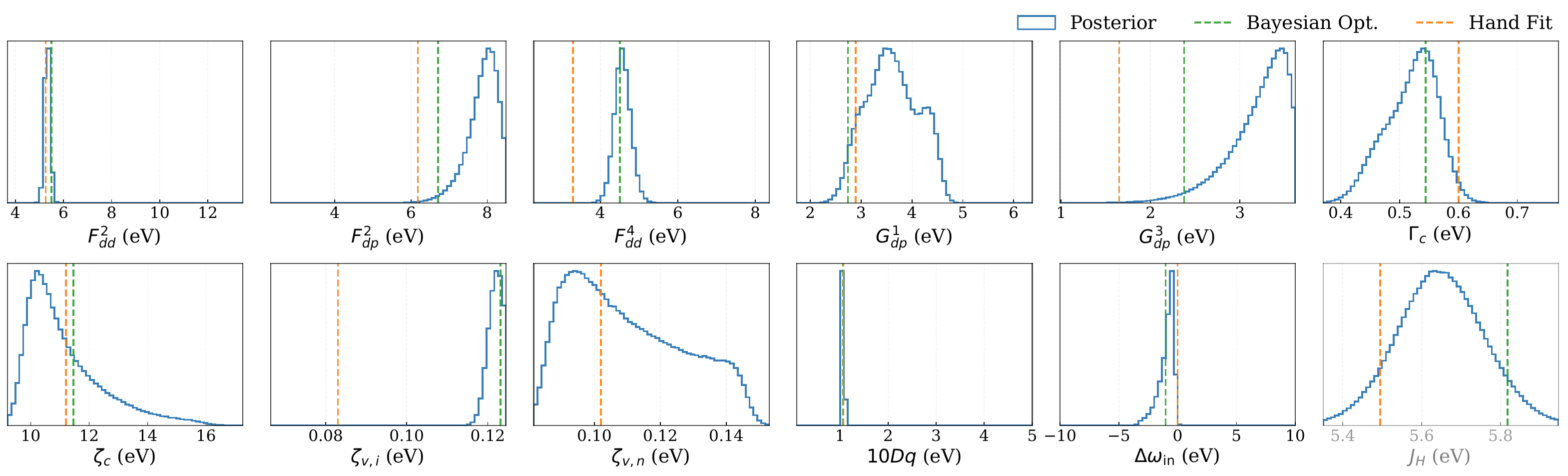}
    \caption{Marginal posterior distributions for all parameters inferred from the observed NiPS$_3$ data. Each panel shows the SBI posterior as a solid blue curve alongside two reference values. The point estimate from \textcite{Lajer2025} is a green dashed line, and a by-hand reference value of \textcite{He2024} is shown as an orange dotted line. The horizontal axis of each panel spans the full prior range, making the degree of posterior concentration directly visible. The Hund's coupling $J_H = F^{(2)}_{dd} + F^{(4)}_{dd}/14$ is not a free parameter of the model but is derived from the posterior samples of the Slater integrals. The Hund's coupling describes the energy separation between spin multiplicities, e.g.\ high-spin and low-spin separations.}
    \label{fig:marginal_comparison}
\end{figure*}

The posterior is substantially concentrated relative to the prior for most parameters as shown in Fig.~\ref{fig:marginal_comparison}, and for the majority of parameters the SBI posterior is consistent with the \textcite{Lajer2025} point estimates. 
This agreement is expected. Where a parameter leaves a sharp imprint on the spectrum, the $L_1$ distance minimized in the reference analysis and the posterior learned by SBI localize it to the same region, and the two methods coincide.

Some parameters are unconstrained by the measurement, for example, the valence spin--orbit coupling $\zeta_{v,n}$ remains close to its prior width, and the core spin--orbit coupling $\zeta_c$ is broad because it is nearly exchangeable with the incident-energy offset $\Delta\omega_{\mathrm{in}}$, a true degeneracy of the forward model examined in the next section. 
To further constrain these parameters additional measurements at other edges, geometries, or energy ranges would be required.
Some parameters like $F^{(2)}_{dd}$ and $F^{(4)}_{dd}$ are tightly constrained by both methods, and for these parameters the posterior width are likely to be limited by single-ion model approximations more than experimental statistics.
The fact that the posterior predictive band is narrow despite the posterior having broad marginals in some parameters demonstrates that the forward model has significant degeneracy that is not captured by point estimates alone.
The slight multimodality visible in some marginals reflects disagreement among the deep ensemble members when applied to experimental data that lies partially outside the distribution of the training simulations. On held-out simulated spectra the ensemble has correct coverage and each member returns a smooth unimodal density, but systematic differences between the single-ion forward model and the real measurement cause individual ensemble members to prefer slightly different regions, producing minor secondary modes in the pooled density. This behavior is precisely why the ensemble is employed---it exposes sensitivity to distributional shift that a single model would mask with overconfident smoothness~\cite{Lakshminarayanan2017,Hermans2022crisis}.

\subsubsection{Parameter correlations}
\label{sec:correlations}

\begin{figure}[htbp]
    \centering
    \includegraphics[width=\linewidth]{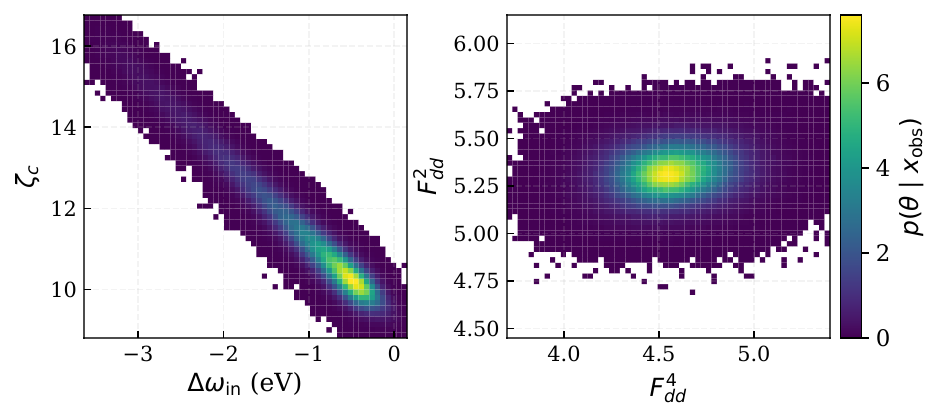}
    \caption{Joint posterior densities for two representative parameter pairs. The intensity encodes the posterior probability density $p(\theta|x_\mathrm{obs})$. \textit{Left:} Core spin-orbit coupling $\zeta_c$ versus incident energy offset $\Delta\omega_{\mathrm{in}}$, showing that the RIXS data constrain a linear combination of these two quantities more tightly than either individually. \textit{Right:} The Slater integrals $F^2_{dd}$ and $F^4_{dd}$, whose joint density shows some correlation and is shifted from the atomic ratio $F^4_{dd}/F^2_{dd}\approx 0.625$.}
    \label{fig:nips3-correlations}
\end{figure}

The joint posterior reveals correlations that are invisible to point estimators. Figure~\ref{fig:nips3-correlations} highlights two representative examples.
For example, the posterior on $(\zeta_c,\,\Delta\omega_{\mathrm{in}})$ shows a clear diagonal correlation.
This occurs because the core spin-orbit coupling and the incident energy offset play similar roles in selecting which intermediate states contribute to the RIXS cross-section, so the data constrain their combination more tightly than either individually.
These correlations imply that an independent measurement of $\zeta_c$ or of $\Delta\omega_{\mathrm{in}}$ would substantially sharpen the complementary marginal without any retraining.
The posterior on $(F^2_{dd},\,F^4_{dd})$ is concentrated along a ridge near the atomic ratio, consistent with the expectation that the measured $dd$ excitation energies determine a linear combination of these parameters rather than either in isolation.

The $(\zeta_c,\,\Delta\omega_{\mathrm{in}})$ degeneracy is also tangible at the level of physical spectra. Forward simulations at five points drawn along the posterior ridge, with core spin-orbit coupling and incident energy offset covarying along the correlation axis, are nearly indistinguishable as shown in Appendix~\ref{app:degeneracy-spectra}.
This is because a larger $\zeta_c$ shifts the energies of intermediate core-hole states in a way that is compensated by a corresponding shift in $\Delta\omega_{\mathrm{in}}$, leaving the observed cross-section approximately unchanged.
This degeneracy is captured naturally by the SBI posterior but would be invisible to any inference method that returns only a point estimate, and resolving it would require an independent constraint on either quantity. More broadly, these correlations reveal that the posterior captures real degeneracies in the space of couplings in a simulation-efficient manner, even though it does not account for all systematic differences between simulation and data. The amortised density estimator learns which parameter combinations the spectra are insensitive to directly from forward simulations, without requiring an explicit analytic model of the degeneracy structure.

\subsubsection{Validation}
\label{sec:validation}

Coverage testing on held-out simulations confirms that the ensemble posterior is statistically consistent with exact inference, and conservative where it deviates. We assess the joint distribution with TARP rather than with per-parameter rank statistics alone, so that poorly estimated correlations are also detected, and we monitor coverage at every round of active learning (Appendix~\ref{app:sbc}). The eigenstate annotations from the SBI posterior samples are in close agreement with those from the reference solution of \textcite{Lajer2025} across nearly all excitations, and the few deviations that do occur are small as shown in Appendix~\ref{app:eigenstate}.
The eigenstate symmetry labels depend nonlinearly on the Hamiltonian parameters, and therefore there is no closed-form map from parameter uncertainty to label uncertainty.
Running the annotation code at each posterior sample turns this into a direct probabilistic statement about the multiplet structure, and any persistent ambiguity in ground-state symmetry would flag a region of parameter space where downstream physical interpretations require additional justification.

\subsubsection{Broadening Sensitivity}
\label{sec:nuisance-validation}

The density estimator is conditioned on $\sigma$, so posteriors can be evaluated at any fixed broadening value without retraining. Examining the conditional posteriors at several fixed $\sigma$ values we find that the posterior has only limited dependence on the assumed broadening. The marginals shift only slightly across the range of $\sigma$ considered, and the marginalized posterior that integrates over $\sigma$ remains consistent with the conditionals at each fixed value. This limited sensitivity constitutes an important physical finding, as it demonstrates that the nuisance parameter is not absorbing explanatory power from the Hamiltonian parameters. If $\sigma$ acted as a compensatory degree of freedom, one would expect the physical parameter posteriors to shift substantially as the broadening is varied, indicating a degeneracy between the energy-loss broadening and the underlying electronic structure. The near-independence observed here confirms that the Hamiltonian parameters are genuinely constrained by the spectral features themselves rather than by an implicit assumption about the broadening.
Full details of the broadening-sensitivity analysis are given in Appendix~\ref{app:nuisance-scan}.

The ability to perform this sensitivity analysis at negligible additional cost is a valuable feature of the amortized approach. Because the network is already trained over the full $\sigma$ range, sweeping over broadening values at inference time requires only re-evaluating the density estimator rather than re-running an expensive sampling procedure. In contrast, iterative approaches such as MCMC require either committing to a single assumed $\sigma$ value or repeating the full inference at each broadening value, making systematic exploration of nuisance sensitivity substantially more expensive.

\subsection{K\texorpdfstring{$_2$}{2}NiF\texorpdfstring{$_4$}{4} data}
\label{sec:k2nif4}

We apply the same TMNRE pipeline---identical encoder architecture, density estimator, and nuisance treatment---to K$_2$NiF$_4$, a layered Ni$^{2+}$ ($3d^8$) compound in a tetragonally distorted octahedral environment. Where the Ni--S bonding in NiPS$_3$ makes it representative of more covalent quantum materials, the ionic Ni--F bonding in K$_2$NiF$_4$ places it much closer to the atomic limit. 
Together with NiPS$_3$, this material probes the method at both ends of the range of bonding character over which the single-ion model is applied.

K$_2$NiF$_4$ crystals were prepared using the flux method \cite{Wanklyn1975flux}. The experimental spectrum was measured at the SIX 2-ID beamline of the NSLS-II at $T=40$~K with an energy resolution of 31~meV full width at half maximum, an incident x-ray angle $\theta_{\mathrm{in}}=10^{\circ}$, a scattering angle $2\Theta=150^{\circ}$, and $\pi$-polarized incident x-rays over an incident-energy range spanning the Ni $L_3$-edge~\cite{klein_2026_21907963}. The measured spectrum was truncated to the energy-loss window $E_{\mathrm{loss}}\in[-0.1,\,4.5]$~eV, yielding a $30\times480$ grid in incident energy and energy loss, respectively.

The K$_2$NiF$_4$ parameterization matches that of NiPS$_3$ so the only differences are in the naming and range of the incident-energy offset, written $\Delta E$ here, and in the bounds placed on the core-hole broadening and the energy-loss broadening, both of which are looser than for NiPS$_3$ as described in Section~\ref{sec:prior}.

Two reference values are compared to the SBI results for K$_2$NiF$_4$. The first is obtained by applying the Bayesian optimization method of \textcite{Lajer2025} to this material. The second, which we refer to as the atomic screened reference, scales the Ni$^{2+}$ $3d^8$ atomic Slater integrals by $0.76$ for the $d$--$d$ terms and $0.87$ for the $d$--$p$ terms, leaves the spin--orbit couplings at their free-ion values, and fixes $10Dq$ at $1$\,eV. 
Reducing the couplings in this way are the standard starting point for atomic multiplet analyses of transition-metal spectra~\cite{Cowan1981,deGroot2008,van1988multiplet}. 
Because the atomic screened reference is fixed in advance of the measurement rather than fitted to it, it indicates where the parameters would be expected to lie for a material near the atomic limit, and the displacement of the posterior away from it is a measure of the additional screening induced by the lattice. 

\begin{figure*}[htbp]
    \centering
    \includegraphics[width=1.\linewidth]{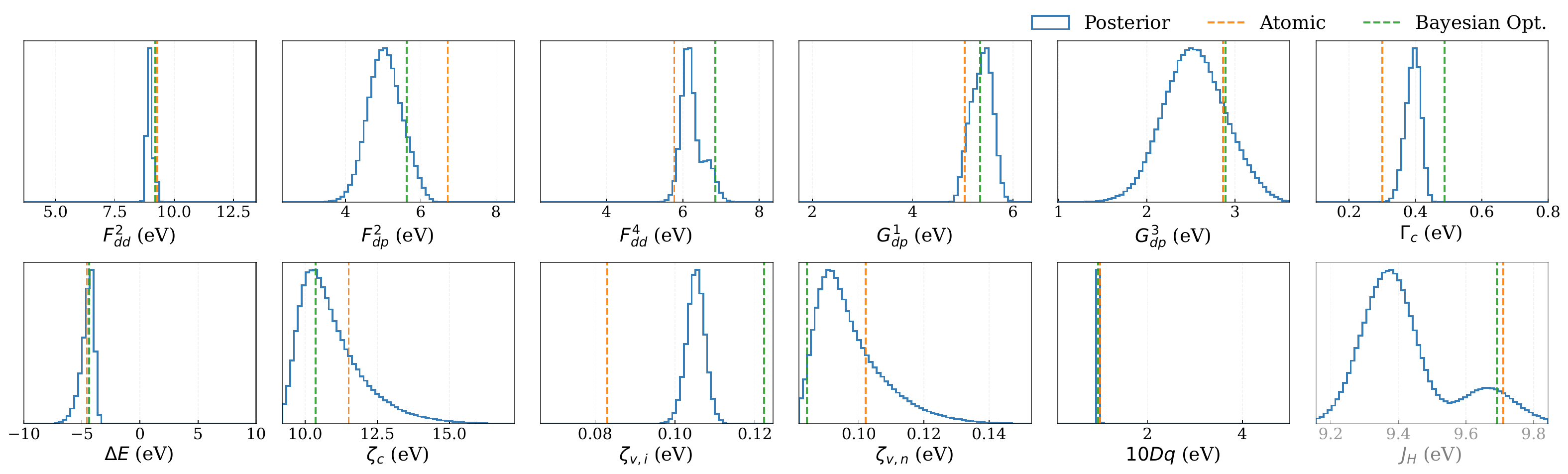}
    \caption{Posterior marginals for K$_2$NiF$_4$ for all 11 interest parameters, obtained with the same TMNRE pipeline used for NiPS$_3$. Each panel shows the SBI posterior as a solid blue curve alongside two reference values. The point estimate from Bayesian optimization of \textcite{Lajer2025} is a green dashed line, and the atomic screened reference value is shown as an orange dotted line. The horizontal axis of each panel spans the full prior range, making the degree of posterior concentration directly visible.}
    \label{fig:k2nif4-marginals}
\end{figure*}

The full spectra of the two reference and the SBI posterior mean all reproduce the main features of the observed spectrum, as shown in Fig.~\ref{fig:k2nif4-spectra}. The posterior mean values for K$_2$NiF$_4$ are also tabulated in Appendix~\ref{app:posterior-mean}.
The screened atomic reference produces a spectrum with some modes at the wrong energy-loss positions, while the Bayesian optimization and SBI posterior mean spectra are both in close agreement with the data.
For most parameters the SBI posterior is consistent with the Bayesian optimization reference, and both are displaced from the atomic screened reference as shown in Fig.~\ref{fig:k2nif4-marginals}.
One exception is the Slater integral $F^4_{dd}$, which favors the atomic screened reference over the Bayesian optimization result while still admitting the Bayesian optimization value within its posterior uncertainty.
As with NiPS$_3$, the SBI posterior exhibits the same characteristic correlation between the core spin--orbit coupling $\zeta_c$ and the incident-energy offset $\Delta E$, reflecting a degeneracy of the forward model that is shared across both materials.

\begin{figure*}[!t]
\centering
\includegraphics[width=1.0\linewidth]{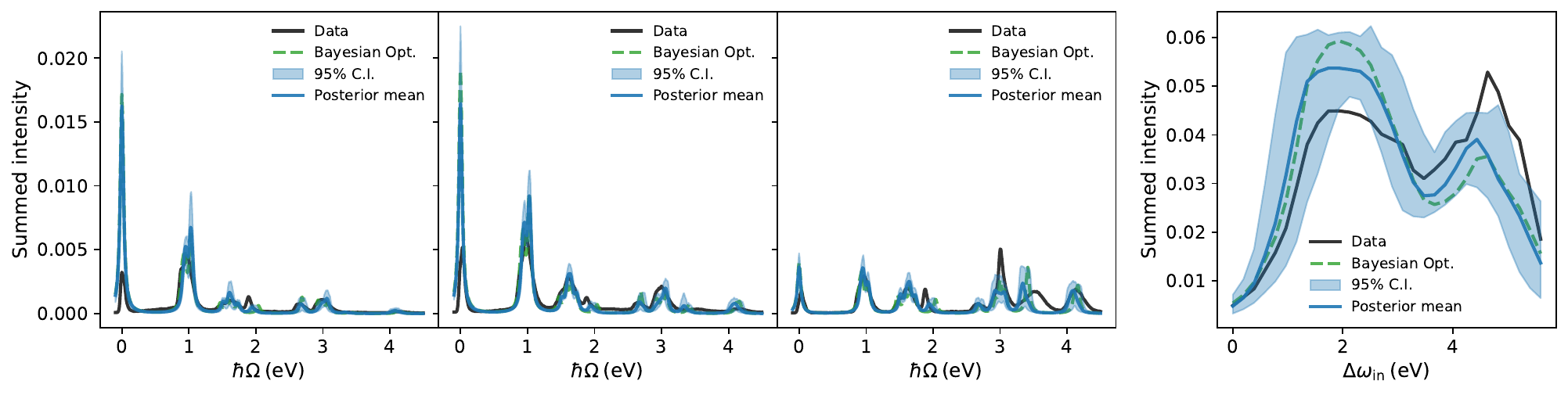}
\caption{Posterior predictive check for K$_2$NiF$_4$. Parameter sets drawn from the posterior are forward-simulated with EDRIXS and projected onto one-dimensional slices. Left panels show three non-overlapping $\Delta\omega_{\mathrm{in}}$ bins of width 4\,eV ($[0,4]$, $[4,8]$, and $[8,12]$\,eV, left to right); the right panel shows summed intensity over all $\hbar\Omega$ as a function of $\Delta\omega_{\mathrm{in}}$. Shaded bands span the 95\% credible interval of the predictive distribution, the solid black curve is the observed data, and the dashed green line is the Bayesian optimization result~\cite{Lajer2025}.}
\label{fig:k2nif4-ppc}
\end{figure*}

\begin{figure}[htbp]
    \centering 
    \includegraphics[width=1.0\linewidth]{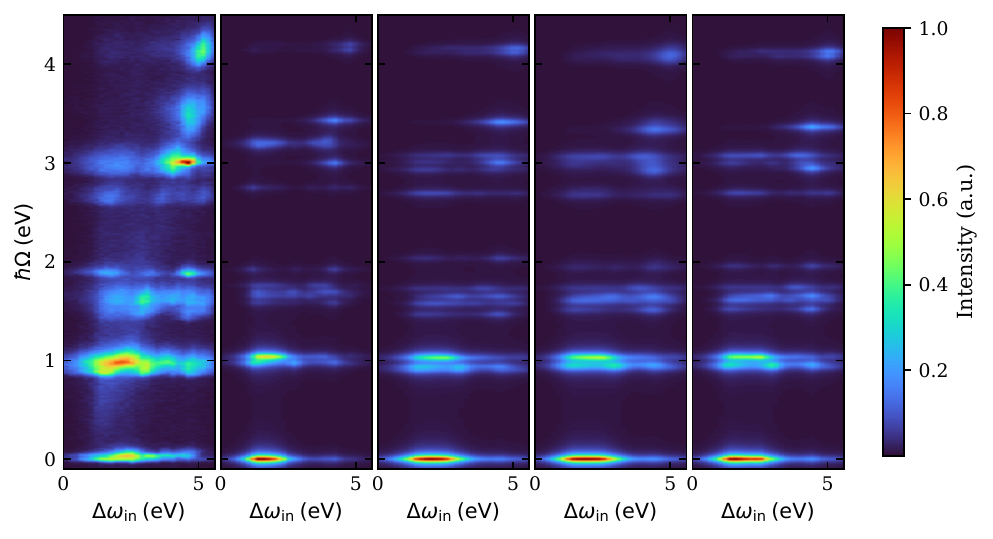}
    \caption{\emph{Far left:} the experimentally measured K$_2$NiF$_4$ RIXS spectrum. \emph{Center left:} the simulated spectrum at the atomic screened reference parameter values. \emph{Center:} the simulated spectrum at the \textcite{Lajer2025} Bayesian optimization reference values. \emph{Center right:} the posterior predictive spectral mean. \emph{Far right:} the simulated spectrum evaluated at the posterior mean of the SBI posterior.}
    \label{fig:k2nif4-spectra}
\end{figure}

The posterior predictive from SBI and the Bayesian optimization reference are both consistent with the observed spectrum in one-dimensional slices in bins of incident energy, as shown in Fig.~\ref{fig:k2nif4-ppc}.
The posterior predictive mean spectrum achieves an $L_1$ distance of $0.608$, comparable to the $0.606$ obtained by Bayesian optimization~\cite{Lajer2025}, while the forward simulation at the posterior mean gives $0.623$ and the best individual posterior sample reaches $0.575$.
However, when integrating over $\hbar\Omega$ to produce the total intensity as a function of incident energy, both posterior predictives fail to faithfully reproduce the observed intensity profiles.
The result of the SBI posterior predictive does produce the two peaks in this spectrum correctly, but the relative intensities are not captured.
Posterior draws are forward-simulated and projected onto one-dimensional slices along the energy-loss axis. The predictive band captures the observed peak positions and relative intensities across all slices, confirming that the inferred posterior is consistent with the measured spectrum. As in the NiPS$_3$ case, the posterior predictive band is narrow, indicating significant degeneracy in the forward model.

\section{Conclusion}
\label{sec:conclusion}

We have presented the first application of simulation-based inference to Hamiltonian parameter extraction from RIXS spectroscopy. 
Using TMNRE with a column-ViT encoder and treating the energy-loss broadening as a nuisance, we recover the full joint posterior over 11 Hamiltonian parameters for NiPS$_3$ and K$_2$NiF$_4$ at a simulation budget comparable to existing optimization approaches. The joint density estimator, trained on simulations drawn only from the marginal-truncated prior with no additional constraints, produces sharp posteriors for both materials, suggesting that a single trained model could serve many compounds sharing the same symmetry class without material-specific prior engineering.
Because SBI targets a posterior under an explicit prior and likelihood rather than implicitly optimizing a norm, the assumptions are auditable and the posterior remains interpretable.

The limitations of our approach include the use of the single-ion EDRIXS model, which neglects charge-transfer physics~\cite{Zaanen1985}, ligand degrees of freedom, and lattice effects, and the Hund's exciton in NiPS$_3$ is a clear case where the forward model systematically fails. 
SBI also introduces additional approximations: the flow matching model is an approximate density estimator, and amortization over the full parameter space may miss fine posterior structure if the simulation budget is insufficient.
We mitigate this with multiple rounds of TMNRE truncation that concentrate simulations in the high-posterior region before training the joint estimator, and by ensembling over independently trained models.
Sum-normalizing spectra before inference means the posterior is conditioned on spectral shape alone, discarding any parameter information carried solely by absolute intensity.
An open question is the noise model in the simulator as we currently assume negligible shot noise, which is defensible if the NiPS$_3$ measurement is high-count, but Poisson augmentation would be required for photon-starved data. 

The amortized SBI posterior opens a class of analyses unavailable to point-estimate methods.
Independent RIXS datasets on the same material can be fused by multiplying posteriors without further simulation. 
The posterior identifies loosely constrained parameters and naturally feeds active experimental design~\cite{sbisupercharged}.
Any quantity derived from $\param$ inherits a proper posterior via predictive sampling. Amortization makes each of these near-free at inference time, and the methodology developed here should generalize to other edges, geometries, and forward models.

\section{Data Availability}
\label{sec:data-availability}

The NiPS$_3$ data that support the findings of this article are available from \textcite{He2024}, and the K$_2$NiF$_4$ data are openly available~\cite{klein_2026_21907963}. The code used to generate the simulations and train and evaluate the models described in this work is openly available at \href{https://github.com/sambklein/sbi-maq-paper}{github.com/sambklein/sbi-maq-paper}. All workflows are managed with Snakemake~\cite{snakemake} and environments with pixi~\cite{pixi}, so that the results in this paper can be regenerated end to end from the raw inputs by a single invocation.

\begin{acknowledgments}
This work was supported by the U.S.\ Department of Energy, Office of Science, Basic Energy Sciences (BES) under the Genesis Mission BES AI Pathfinder Program, MAIQMag: Multimodal AI for 2D Quantum Magnets. M.K. is supported by the US Department of Energy (DOE) under Grant No. DE-AC02-76SF00515.  V.B.~was supported by the U.S. Department of Energy, Office of Science, Office of Basic Energy Sciences,
under Award No.~DE-SC0012704. W.H.\ acknowledges the support by the U.S.\ Department of Energy, Office of Science, Basic Energy Sciences, Materials Sciences and Engineering Division, under contract DE-AC02-76SF00515.
 This research used resources at the SIX beamline of the NSLS-II, a U.S. DOE Office of Science User Facility operated for the DOE Office of Science by Brookhaven National Laboratory under Contract No.~DESC0012704.
\end{acknowledgments} 

\clearpage
\appendix

\section{Encoder Architecture Ablations}
\label{app:encoder-ablation}

\subsection{Full design grid}

All candidates are trained on the 100{,}000 restricted-prior simulations described in Section~\ref{sec:encoder-selection} with the training configuration of Appendix~\ref{app:flow-matching-details}, and evaluated on a held-out set of 5{,}000 further simulations drawn from the same restricted prior. TARP credibility levels are computed with one randomly drawn reference point per test spectrum and 1{,}000 posterior samples per test spectrum, with all parameters normalized to the unit hypercube before computing distances. Sharpness is the per-parameter 90\% credible width normalized by the width of the restricted prior and averaged over the 11 interest parameters. The candidates span five architecture classes. The PCA baseline projects each spectrum onto the components retaining 95\% of the variance and passes the scores through an MLP, discarding the spatial layout entirely; the flat baseline passes raw flattened pixels through an MLP; the convolutional baseline applies a depthwise-separable backbone followed by global pooling; the hybrid applies the same backbone but tokenizes its output by column before a transformer head; and the column and row Vision Transformers tokenize the raw map along the incident-energy and energy-loss axes respectively. Table~\ref{tab:encoder-comparison-full} reports all configurations.

The ordering is driven by how much of the physical layout each architecture preserves. The PCA baseline achieves the lowest TARP deviation (0.012) but returns a posterior barely narrower than the restricted prior (sharpness 0.45), which illustrates why coverage cannot be used as a selection criterion on its own. The flat MLP, CNN, and row-ViT baselines have moderate coverage (0.039--0.077) and are uninformative (sharpness 0.34--0.38). The column-ViT architectures are conservatively covered (0.057--0.066) and dramatically sharper than every other encoder (0.19--0.25), with the gain attributable to the tokenization rather than to local feature extraction, since applying a convolutional backbone before column tokenization performs comparably to using raw pixels. We note that the flat MLP baseline was given an encoder weight decay of $0.1$ rather than the $0.01$ used for every other architecture, as it overfits severely otherwise, so it is not matched to the others in regularization.

\begin{table}[t]
\caption{Full encoder architecture comparison. TARP dev.\ is the mean absolute deviation of the TARP credibility distribution from uniformity, a measure of coverage, and sharpness is the 90\% credible width averaged over parameters and normalized by the restricted prior width, with lower better for both. All models are trained on the same 100{,}000 restricted-prior simulations and evaluated on the same 5{,}000 held-out simulations, and all use dropout 0.2. The selected architecture is highlighted in bold.}
\label{tab:encoder-comparison-full}
\centering
\begin{tabular}{@{}lrr@{}}
\toprule
Encoder & TARP dev. & Sharpness \\
\midrule
PCA $\to$ MLP                      & $0.012$ & $0.45$ \\
flatten $\to$ MLP                  & $0.077$ & $0.34$ \\
CNN $\to$ global pool              & $0.039$ & $0.34$ \\
row-ViT ($d{=}128$, 4 layers)      & $0.077$ & $0.38$ \\
CNN $\to$ col-ViT                  & $0.059$ & $0.35$ \\
col-ViT ($d{=}64$, 4 layers)       & $0.057$ & $0.19$ \\
\textbf{col-ViT ($d{=}128$, 4 layers)}  & $\mathbf{0.060}$ & $\mathbf{0.22}$ \\
col-ViT ($d{=}128$, 2 layers)      & $0.066$ & $0.19$ \\
col-ViT ($d{=}128$, 2 layers, narrow) & $0.066$ & $0.25$ \\
\bottomrule
\end{tabular}
\end{table}

\subsection{Ensembling across architectures}

Coverage measured on a single model need not carry over to the ensemble that is actually deployed. Pooling several independently initialized models can rescue an architecture whose individual posteriors are poorly covered, so the ordering in Table~\ref{tab:encoder-comparison-full} could in principle change once each architecture is ensembled, and the only way to establish whether it does is to build the ensembles and measure them. We therefore trained five independently initialized models for the selected column-ViT and for the flat and PCA baselines and computed the coverage of each pooled posterior. The column-ViT ensemble achieves a TARP deviation of 0.054, improving slightly on the single-model value of 0.060, while the PCA and flat baselines remain at 0.014 and 0.078 respectively, close to their single-model values of 0.012 and 0.077. The ordering is unchanged: the column-ViT ensemble remains conservatively covered and is dramatically sharper than the baselines, so the architecture selected on single-model performance is also the one to select for the deployed ensemble.

\section{Flow Matching Training Details}
\label{app:flow-matching-details}

The encoder and velocity network are trained jointly for up to 300 epochs with early stopping on a held-out validation set. We use AdamW~\cite{Loshchilov2019} with a learning rate of $10^{-3}$ and batch size 256, with a differential learning rate for the encoder (0.1$\times$) that warms up linearly over the first 25\% of training. An exponential moving average of the velocity network weights with decay 0.999 is maintained and used for validation scoring and final sampling. Each member of the deep ensemble is trained independently with a different random initialization and data shuffle order. Training a single flow matching model takes approximately 43 minutes for K$_2$NiF$_4$ and 31 minutes for NiPS$_3$ on a GeForce RTX 2080 Ti GPU. The final TMNRE classifier (round 7) trains in approximately 3.5 minutes for K$_2$NiF$_4$ and 8 minutes for NiPS$_3$ on the same hardware.

\section{Summary Statistics for TMNRE Rounds}
\label{app:summaries}

During the TMNRE truncation rounds, the full 2D RIXS spectrum is compressed into a low-dimensional summary to enable efficient classifier training on small datasets. We flatten each spectrum and apply PCA, retaining components up to 95\% cumulative explained variance with a maximum of 64 components. These PCA scores are concatenated with the first two moments of the energy-loss profile computed in 10 equally spaced slices along the incident-energy axis. The intensity-weighted centroid of each slice locates the excitations in energy loss and the corresponding width measures their broadening, providing complementary information about peak positions and shapes that PCA alone may distribute across many components. The combined summary vector is standardized to zero mean and unit variance before being passed to the NRE classifiers. This summary is used only for the TMNRE rounds; the final joint density estimator operates on the full spectrum via the column-ViT encoder.

For K$_2$NiF$_4$, the energy-loss axis is cropped below 0.5\,eV before computing any summary statistics, removing the elastic line. The elastic peak is dominated by the instrumental broadening function rather than by the Hamiltonian parameters of interest, so retaining it would waste PCA variance capacity on an uninformative feature and bias the energy-loss moments toward the zero-loss region. The cutoff is applied identically to both the simulated training spectra and the experimental observation, and all subsequent operations---normalization, PCA, and moment computation---use only the inelastic portion of the spectrum. Because this summary is used only during the TMNRE truncation rounds, the elastic-line crop does not affect the joint density estimator, which receives the full uncropped spectrum through the column-ViT encoder. This preprocessing is not needed for NiPS$_3$, whose energy-loss grid already begins above the elastic line.

\section{TMNRE Classifier Architecture}
\label{app:tmnre-classifier}

Each truncation round trains a batched ensemble of binary classifiers to estimate the marginal log-ratio $\log r(\theta_d \mid x) = \log p(\theta_d \mid x) - \log p(\theta_d)$ for every parameter simultaneously. The architecture is a two-hidden-layer MLP with 64 units per layer and ReLU activations, taking as input the concatenation of a single standardized parameter value and the summary vector described in Appendix~\ref{app:summaries}. Five independently initialized copies are trained for each parameter and their log-ratio outputs are averaged, giving $5 \times 12 = 60$ classifiers evaluated in a single batched forward pass via batched matrix multiplications. Training uses AdamW with learning rate $10^{-3}$ and no weight decay for 500 epochs on all accumulated simulations up to that round, with 10 fresh negative permutations drawn each epoch to prevent memorization of specific joint-vs-marginal pairings. Log-ratios are evaluated on a two-pass adaptive grid of 2500 points per parameter, first a coarse sweep across the full prior and then a refined sweep around the detected peak region, and truncation retains the interval where the normalized ratio exceeds $\varepsilon = 10^{-6}$. An empirical coverage test on 200 held-out samples confirms that the resulting HPD intervals contain the true value at close to their nominal rate at each round.

\section{Prior Bounds}
\label{app:prior-bounds}

Table~\ref{tab:prior-bounds} lists the prior bounds for all parameters in both materials. All priors use a Tukey-windowed (raised-cosine taper) smooth-box form with taper fraction 0.1. The Slater integrals are scaled relative to Ni$^{2+}$ $3d^8$ atomic reference values from the EDRIXS database, and $F^{(4)}_{dd}$ is drawn as a ratio $r \sim \mathcal{N}(0.625,\,0.1)$ applied to $F^{(2)}_{dd}$ rather than sampled independently.

\begin{table}[t]
\caption{Prior bounds for NiPS$_3$ and K$_2$NiF$_4$. Slater integrals are expressed as scale factors applied to atomic reference values. The energy-loss broadening $\sigma$ is treated as a nuisance parameter in both materials.}
\label{tab:prior-bounds}
\centering
\begin{tabular}{@{}lcc@{}}
\toprule
Parameter & NiPS$_3$ & K$_2$NiF$_4$ \\
\midrule
$F^{(2)}_{dd}$ & $[0.3,\,1.1]\times P_\mathrm{at}$ & $[0.3,\,1.1]\times P_\mathrm{at}$ \\
$F^{(4)}_{dd}$ & $r \cdot F^{(2)}_{dd}$ & $r \cdot F^{(2)}_{dd}$ \\
$F^{(2)}_{dp}$ & $[0.3,\,1.1]\times P_\mathrm{at}$ & $[0.3,\,1.1]\times P_\mathrm{at}$ \\
$G^{(1)}_{dp}$ & $[0.3,\,1.1]\times P_\mathrm{at}$ & $[0.3,\,1.1]\times P_\mathrm{at}$ \\
$G^{(3)}_{dp}$ & $[0.3,\,1.1]\times P_\mathrm{at}$ & $[0.3,\,1.1]\times P_\mathrm{at}$ \\
$\zeta_c$ & $[0.8,\,1.5]\times\zeta_\mathrm{at}$ & $[0.8,\,1.5]\times\zeta_\mathrm{at}$ \\
$\zeta_{v,i}$ & $[0.8,\,1.5]\times\zeta_\mathrm{at}$ & $[0.8,\,1.5]\times\zeta_\mathrm{at}$ \\
$\zeta_{v,n}$ & $[0.8,\,1.5]\times\zeta_\mathrm{at}$ & $[0.8,\,1.5]\times\zeta_\mathrm{at}$ \\
$\Gamma_c$ (eV) & $[0.37,\,0.77]$ & $[0.1,\,0.8]$ \\
$\sigma$ (eV) & $[0.030,\,0.036]$ & $[0.01,\,0.05]$ \\
$10Dq$ (eV) & $[0.1,\,5.0]$ & $[0.1,\,5.0]$ \\
$\Delta\omega_\mathrm{in}$ / $\Delta E$ (eV) & $[-10,\,10]$ & $[-10,\,10]$ \\
\bottomrule
\end{tabular}
\end{table}

\section{NiPS$_3$ Full Pairplot}
\label{app:pairplot}

\begin{figure*}[htbp]
    \centering
    \includegraphics[width=\linewidth]{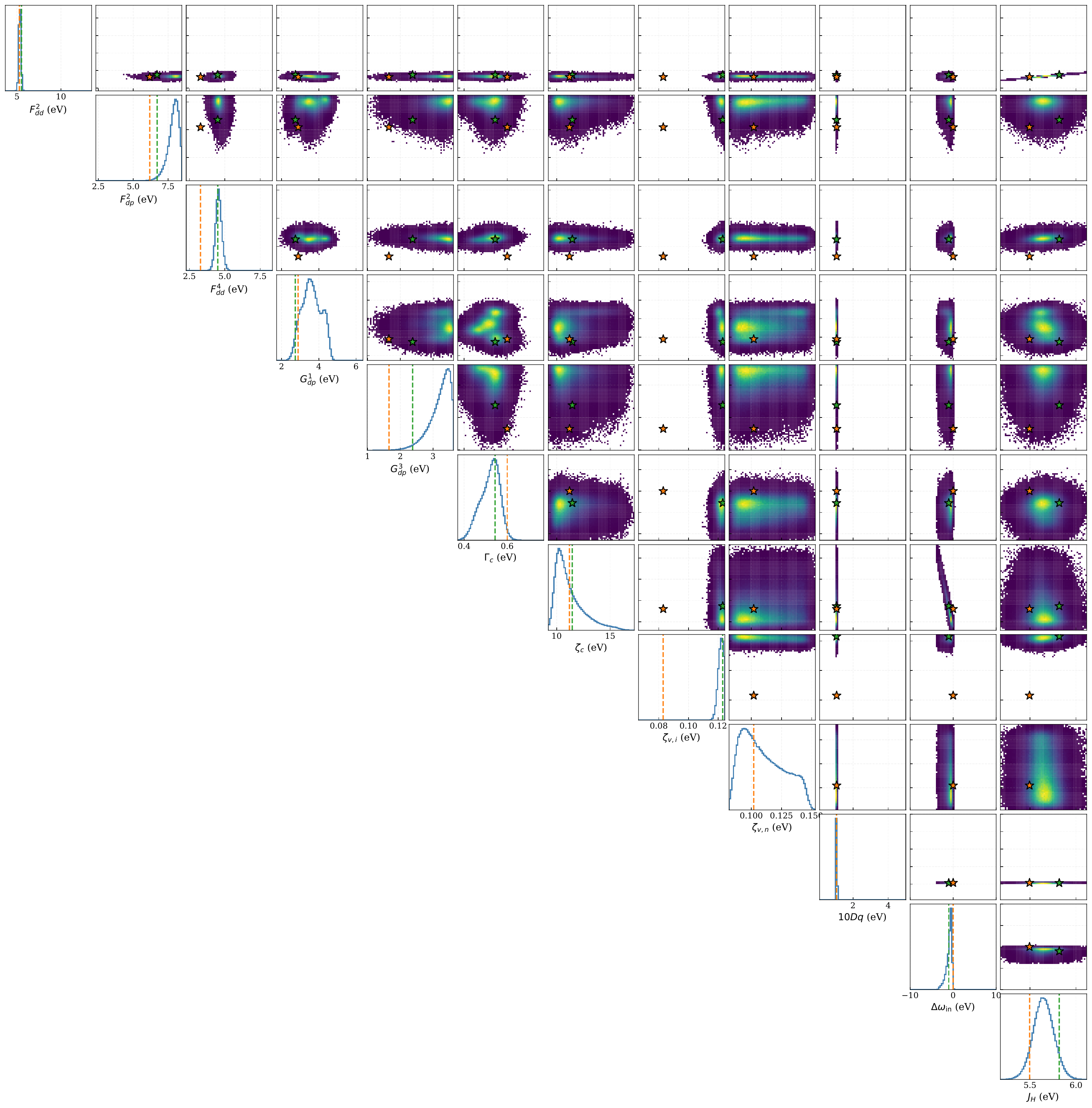}
    \caption{Full posterior for NiPS$_3$ over the 11 Hamiltonian parameters together with the derived Hund's coupling $J_H$. The diagonal shows the 1D marginals and the off-diagonal panels show the 2D joint posteriors. The axes are zoomed to the support of the posterior, and the result from \textcite{Lajer2025} is overlaid as a vertical green dashed line and the result from \textcite{He2024} is overlaid as a vertical orange line.}
    \label{fig:nips3-full-pairplot}
\end{figure*}

Figure~\ref{fig:nips3-full-pairplot} shows the complete NiPS$_3$ posterior over the 11 Hamiltonian parameters together with the derived Hund's coupling $J_H$, with all 1D marginals and 2D joint posteriors. Whereas the marginal summaries in Fig.~\ref{fig:marginal_comparison} span the full prior range to make the degree of posterior concentration visible, here the axes are zoomed to the support of the posterior itself. This resolves the fine structure of the joint distributions and exposes correlations between parameters that are otherwise compressed against the prior bounds, of which the panels in Fig.~\ref{fig:nips3-correlations} are representative examples. No reference values are overlaid, so that the figure reflects only what the data constrain. The full pairplot is included here for completeness; selected panels are discussed in Section~\ref{sec:nips3}.

\section{K\texorpdfstring{$_2$}{2}NiF\texorpdfstring{$_4$}{4} Full Pairplot}
\label{app:k2nif4}

\begin{figure*}[htbp]
    \centering
    \includegraphics[width=\linewidth]{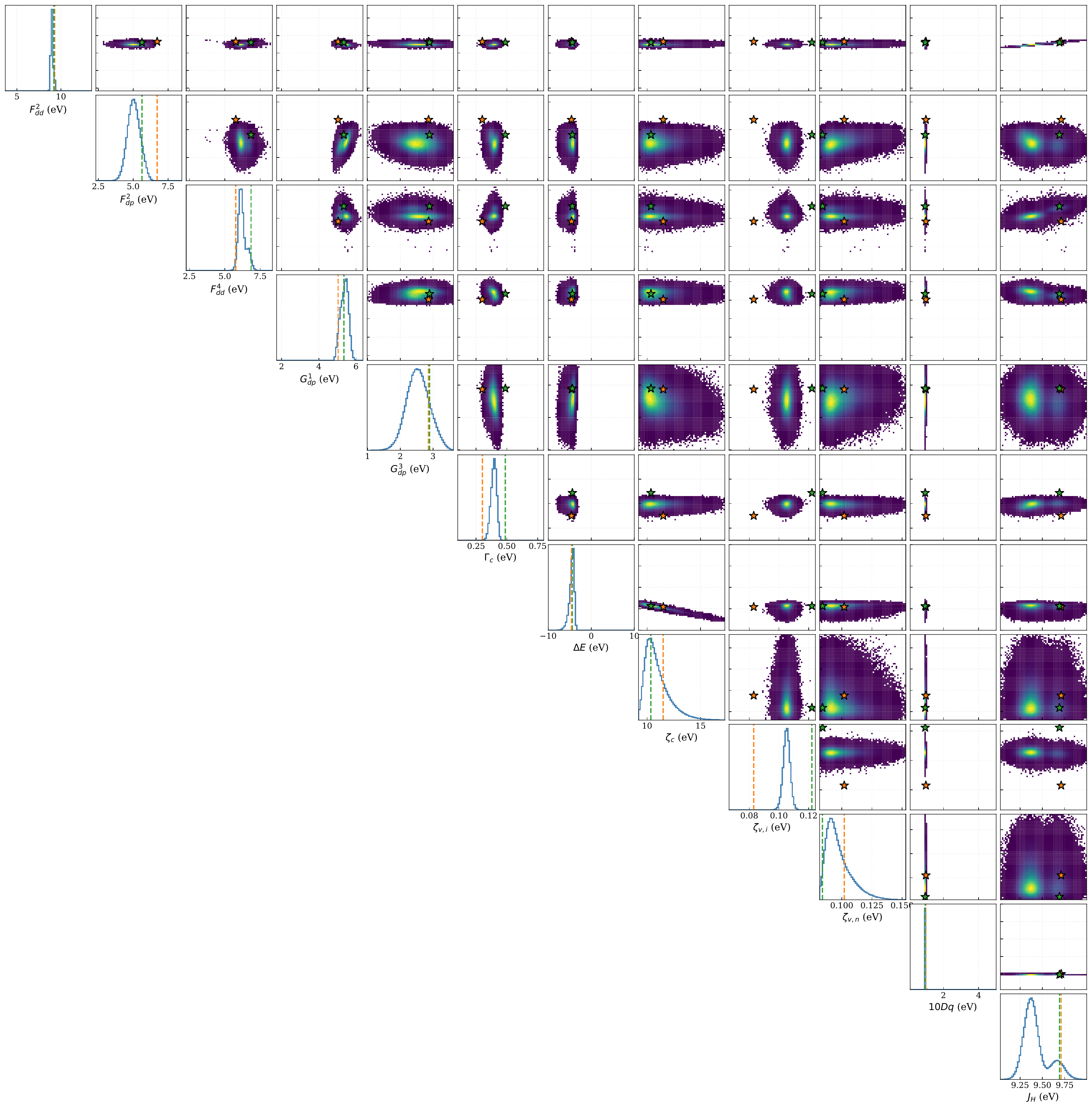}
    \caption{Full posterior for K$_2$NiF$_4$ over all 11 interest parameters. The diagonal shows the 1D marginals and the off-diagonal panels show the 2D joint posteriors. The axes are zoomed to the support of the posterior, and the result from \textcite{Lajer2025} is overlaid as a vertical green dashed line and the atomic screened reference is overlaid as a vertical orange line.}
    \label{fig:k2nif4-full-pairplot}
\end{figure*}

Figure~\ref{fig:k2nif4-full-pairplot} shows the complete K$_2$NiF$_4$ posterior over all 11 interest parameters, with all 1D marginals and 2D joint posteriors. Whereas the marginal summaries in Fig.~\ref{fig:k2nif4-marginals} span the full prior range to make the degree of posterior concentration visible, here the axes are zoomed to the support of the posterior itself. This resolves the fine structure of the joint distributions and exposes correlations between parameters that are otherwise compressed against the prior bounds. The Bayesian optimization result of \textcite{Lajer2025} and the atomic screened reference are overlaid as in the main-body marginals. The full pairplot is included here for completeness; the 1D marginals are discussed in Section~\ref{sec:k2nif4}.

\section{Spectral Degeneracy Along the \texorpdfstring{$(\zeta_c,\,\Delta\omega_{\mathrm{in}})$}{(zeta\_c, delta omega\_in)} Ridge}
\label{app:degeneracy-spectra}

\begin{figure}[htbp]
    \centering
    \includegraphics[width=\linewidth]{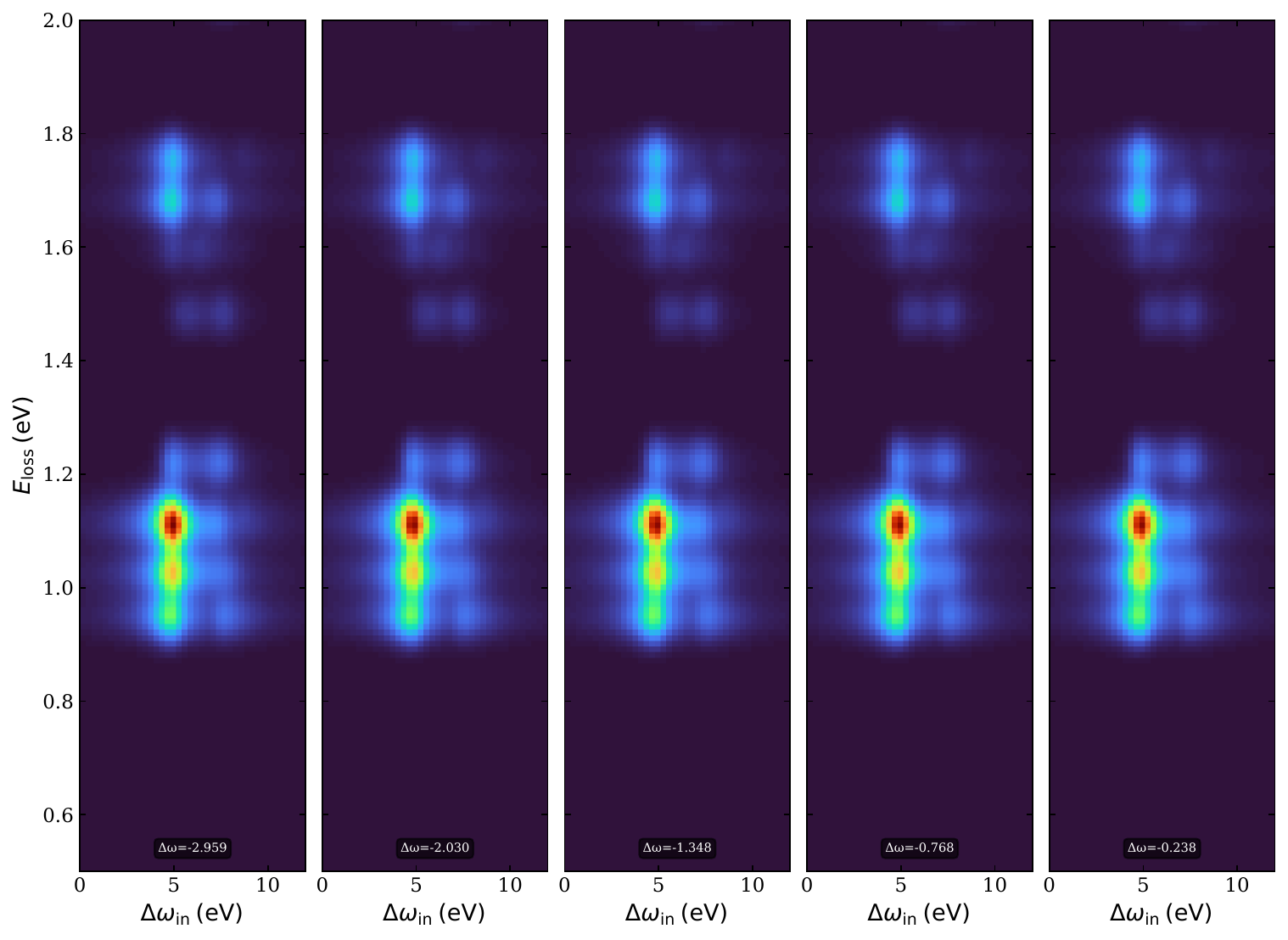}
    \caption{Simulated RIXS spectra at five parameter combinations drawn along the posterior ridge in $(\zeta_c,\,\Delta\omega_{\mathrm{in}})$ space (cf.\ Figure~\ref{fig:nips3-correlations}, left panel). Each column corresponds to one point on the degeneracy axis, with $\zeta_c$ and $\Delta\omega_{\mathrm{in}}$ covarying to maintain approximate agreement with the data. The spectra are nearly indistinguishable across the five samples, confirming that the two parameters are nearly exchangeable over the posterior support: their combination, rather than either individually, is what the RIXS measurement constrains.}
    \label{fig:correlation_panel}
\end{figure}

The NiPS$_3$ posterior exhibits a pronounced correlation between the core-hole spin--orbit coupling $\zeta_c$ and the incident-energy detuning $\Delta\omega_{\mathrm{in}}$, which manifests as an elongated ridge in their joint distribution. Such a ridge signals a near-degeneracy in the forward model where the data constrain a combination of the two parameters far more tightly than either one alone. To confirm that this correlation reflects a genuine spectral degeneracy rather than an artifact of the inference, we sample five $(\zeta_c,\,\Delta\omega_{\mathrm{in}})$ pairs spaced along the ridge and simulate the corresponding RIXS spectra directly with EDRIXS. Figure~\ref{fig:correlation_panel} shows the result: the spectra are nearly indistinguishable across all five samples, so the two parameters trade off against one another almost exactly over the posterior support. The measurement therefore identifies their joint combination, and the marginal width of each parameter individually reflects this exchangeability rather than a lack of sensitivity in the data.

\section{Posterior Coverage}
\label{app:sbc}

Rank-based diagnostics test whether an approximate posterior is statistically consistent with exact inference. The general procedure draws parameter sets $\param^{(i)}$ from the prior, simulates a synthetic observation $x^{(i)}$ at each with EDRIXS, runs inference to obtain a posterior $q(\param \mid x^{(i)})$, draws samples from it, and records where the true value falls within those samples. If the approximate posterior is exact, the resulting ranks are uniformly distributed, and any departure from uniformity identifies a specific failure mode. Ranks that accumulate at both extremes indicate an under-dispersed, overconfident posterior, ranks that concentrate in the middle indicate an over-dispersed posterior, and ranks that drift toward one end indicate a systematic bias. Applied to each parameter separately this is simulation-based calibration~\cite{Talts2018}, which certifies the one-dimensional marginals but is insensitive to misestimated correlations in the joint distribution.

We therefore evaluate the ensemble with TARP~\cite{Lemos2023}, introduced in Section~\ref{sec:encoder-selection}, which probes the joint posterior directly by measuring how often the true parameter vector lies closer to a randomly drawn reference point than the posterior samples do. The held-out test set consists of $5{,}000$ parameter sets drawn from the TMNRE-restricted prior and simulated with EDRIXS, none of which are seen during training. For each test spectrum we draw $1{,}000$ posterior samples from each of the five ensemble members and pool them, giving $5{,}000$ samples per test point, and we compute TARP credibility levels with one randomly drawn reference point per test spectrum and all parameters normalized to the unit hypercube before computing distances. We report the empirical CDF of the resulting credibility levels, where a posterior with perfect coverage traces the diagonal and a conservative posterior wider than the truth bows above it. Alongside the joint test we retain the per-parameter marginal rank statistics computed from the same pooled samples, which localize any coverage failure to particular Hamiltonian parameters.

The active learning rounds are monitored with a cheaper diagnostic. At each round the marginal ratio estimators are used to construct highest-posterior-density intervals at a range of nominal credible levels for a sample of parameter sets, and we check that the intervals contain the true value at close to the nominal rate. This confirms that the truncation threshold is not discarding regions the data support, though it is computed from the same simulations used to fit the classifiers and so is a consistency check rather than an independent test.

Because the flow matching model is trained on data drawn from the TMNRE-restricted prior rather than the full prior, the learned posterior cannot be overconfident in regions outside the restricted support. Probability mass in those regions is not penalized during training, so the model tends to place excess mass in the tails relative to the true posterior. This conservative bias is a known and expected property of sequential methods and is preferable to overconfidence, and it is the direction in which we expect any deviation from the diagonal to appear.

\section{Sources of Uncertainty in the Reported Posterior}
\label{app:ensemble-uncertainty}

Model uncertainty in the reported posteriors is captured by the deep ensemble. Five flow matching models are trained independently on the same restricted-prior dataset, differing only in random initialization and data shuffling, and their posterior samples are pooled. Any disagreement between ensemble members about the shape or location of the posterior therefore broadens the pooled distribution, which is the behaviour we want from a diagnostic for overconfidence in a single approximate model. The TMNRE truncation rounds do not use the encoder at all; they operate directly on the low-dimensional summaries of Appendix~\ref{app:summaries}, so this discussion applies only to the final joint density estimation stage.

\section{Nuisance Parameter Sensitivity: Full Marginals}
\label{app:nuisance-scan}
\begin{figure*}
\centering
\includegraphics[width=\textwidth]{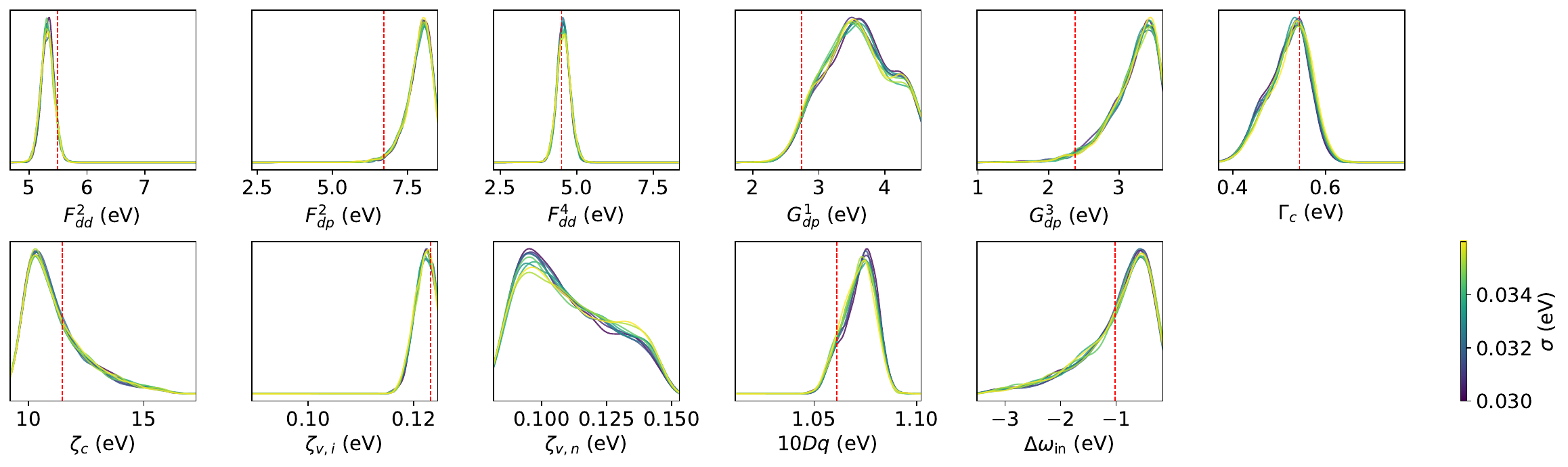}
\caption{Posterior marginals for all 11 inferred Hamiltonian parameters conditioned on several fixed values of the energy-loss broadening $\sigma$. Each curve is obtained by evaluating the amortized network $q_\phi(\param \mid \obs, \sigma)$ at a fixed $\sigma$ without retraining. $\Gamma_c$ and $\zeta_{v,i}$ shift appreciably as $\sigma$ varies, while the remaining parameters are largely insensitive, indicating that the bias from fixing $\sigma$ to a point estimate is concentrated in those two quantities.}
\label{fig:nuisance-scan-all}
\end{figure*}

Figure~\ref{fig:nuisance-scan-all} displays the posterior marginals $q_\phi(\theta_i \mid \obs, \sigma)$ evaluated at several fixed values of the energy-loss broadening $\sigma$ on the NiPS$_3$ posterior. There is very little dependence on $\sigma$ for most parameters, consistent with the small prior width assigned to $\sigma$ for this material.

\begin{figure*}
\centering
\includegraphics[width=\textwidth]{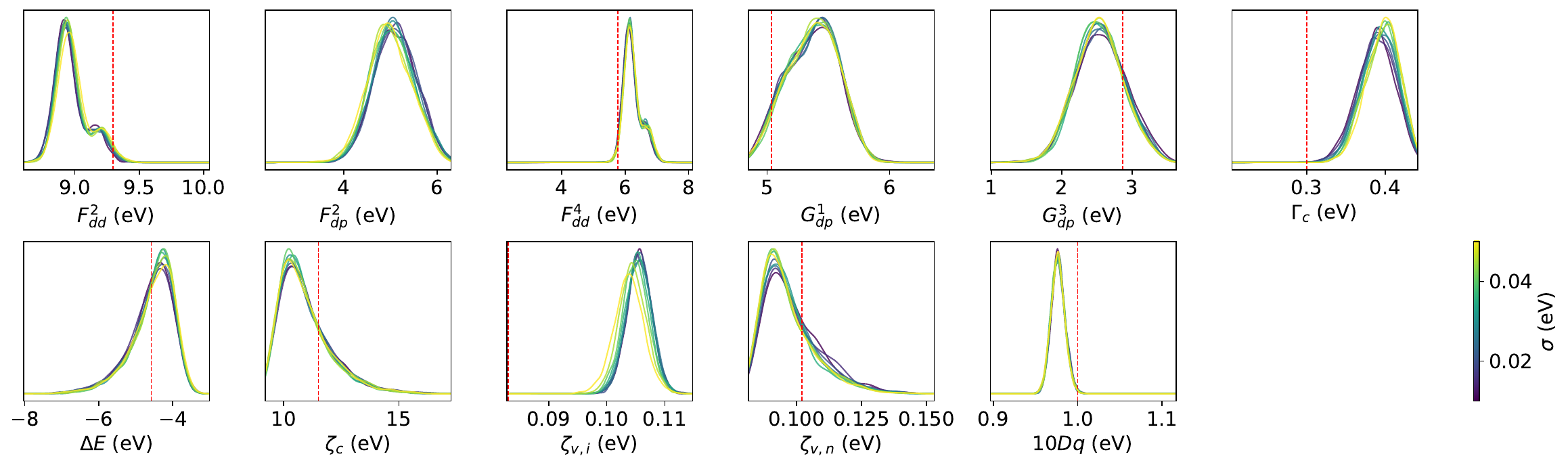}
\caption{Broadening sensitivity for K$_2$NiF$_4$: posterior marginals for all 11 inferred Hamiltonian parameters conditioned on several fixed values of the energy-loss broadening $\sigma$. As for NiPS$_3$ (Fig.~\ref{fig:nuisance-scan-all}), most parameters are largely insensitive to the assumed broadening despite the broader $\sigma$ prior used for this material.}
\label{fig:nuisance-scan-k2nif4}
\end{figure*}

Figure~\ref{fig:nuisance-scan-k2nif4} shows the analogous broadening-sensitivity analysis for K$_2$NiF$_4$. Despite the broader prior on $\sigma$ used for this material ($[0.01,\,0.05]$\,eV versus $[0.030,\,0.036]$\,eV for NiPS$_3$), the conditional posteriors again exhibit only modest dependence on the assumed broadening for most parameters, confirming that the Hamiltonian parameters are constrained by spectral features rather than by an implicit broadening assumption. 
There are some variations in $\Gamma_c$ and $\zeta_{v,i}$, but they are relatively minor compared to width of the marginal posteriors.

\section{Eigenstate Annotation}
\label{app:eigenstate}

\begin{figure}[htbp]
    \centering
    \includegraphics[width=\columnwidth]{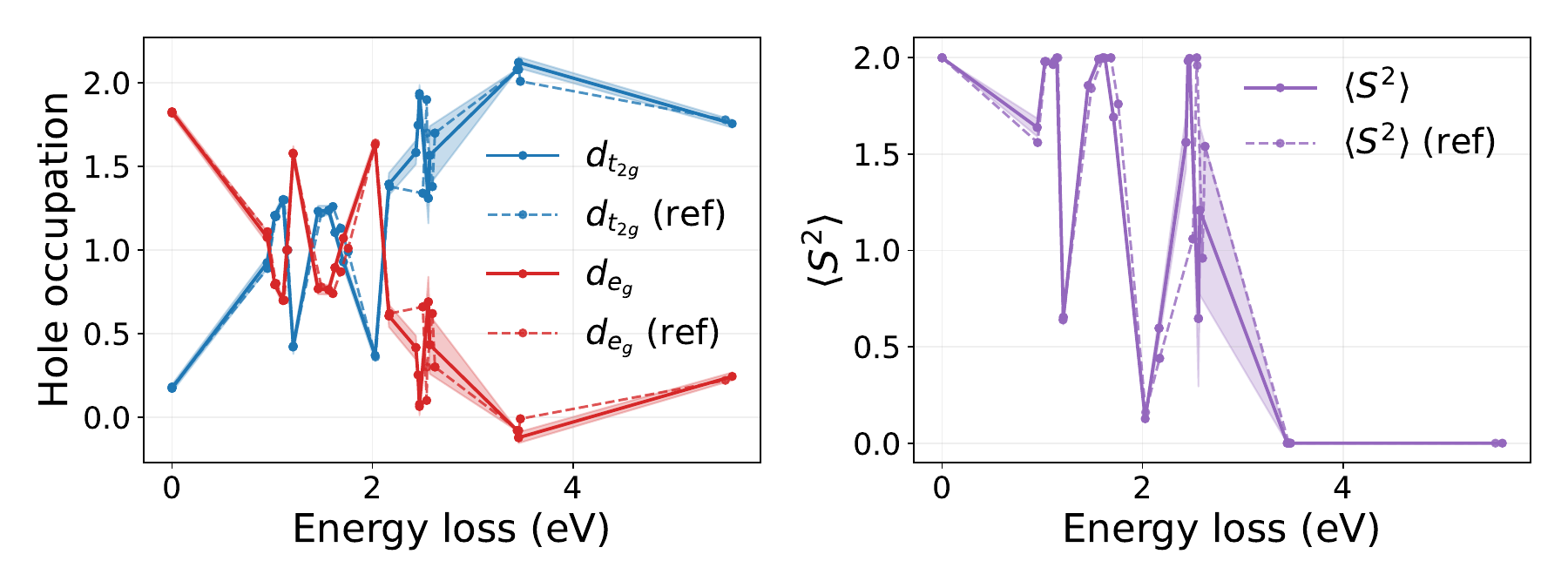}
    \caption{Eigenstate annotation for NiPS$_3$ at the posterior mean and across posterior samples. The SBI solutions agree closely with the \textcite{Lajer2025} reference annotations; small deviations are within posterior uncertainty.}
    \label{fig:nips3_eigenstate_annotation}
\end{figure}

The symmetry labels of eigenstates depend nonlinearly on the Hamiltonian parameters, so there is no closed-form propagation of parameter uncertainty to label uncertainty. We instead evaluate the eigenstate annotation code of \textcite{Lajer2025} at the posterior mean returned by SBI and at a set of posterior draws, thereby mapping the inferred parameter distribution onto a distribution over symmetry labels. Figure~\ref{fig:nips3_eigenstate_annotation} shows the resulting annotations for NiPS$_3$. The posterior mean labels agree with the reference annotations of \textcite{Lajer2025} across nearly all excitations, and the few deviations that do occur fall within the spread of labels obtained across posterior samples. This agreement provides an independent consistency check: because the eigenstate labeling was not part of the training objective, the concordance confirms that the inferred posterior concentrates in a physically meaningful region of parameter space rather than merely reproducing the spectral data through a degenerate combination of parameters.

\begin{figure}[htbp]
    \centering
    \includegraphics[width=\columnwidth]{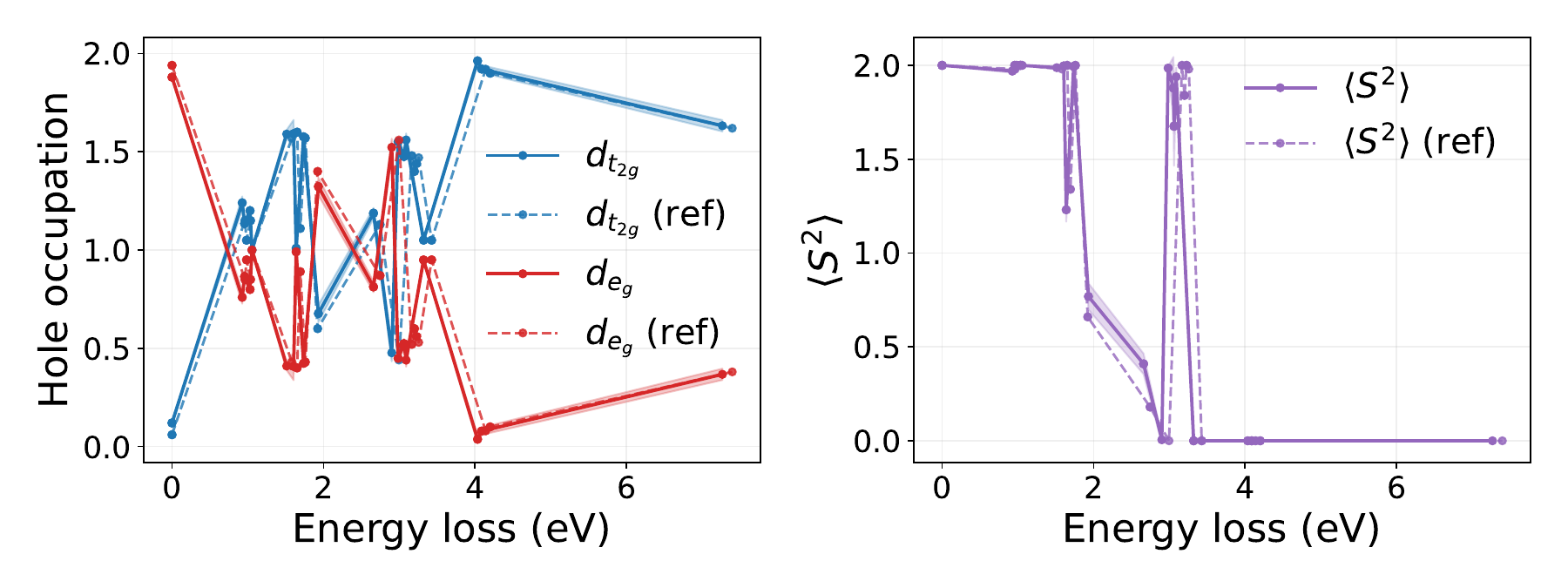}
    \caption{Eigenstate annotation for K$_2$NiF$_4$ at the posterior mean and across posterior samples. As for NiPS$_3$, the SBI solutions agree with the reference annotations; deviations lie within posterior uncertainty.}
    \label{fig:k2nif4_eigenstate_annotation}
\end{figure}

Figure~\ref{fig:k2nif4_eigenstate_annotation} shows the analogous eigenstate annotation for K$_2$NiF$_4$. The same procedure is applied: we evaluate symmetry labels at the SBI posterior mean and across posterior draws, and compare against the reference solution. As in the NiPS$_3$ case, the posterior mean labels are consistent with the reference annotations, and the variation across posterior samples captures the residual uncertainty in the symmetry assignments.

\section{The $p(\sigma \mid x) = p(\sigma)$ Assumption}
\label{app:sigma-assumption}

The marginalized posterior we report is
\begin{equation*}
p(\param \mid \obs) = \int q_\phi(\param \mid \obs, \sigma)\,p(\sigma)\,d\sigma.
\end{equation*}
This differs from the true Bayesian marginal
\begin{equation*}
p(\param \mid \obs) = \int p(\param \mid \obs, \sigma)\,p(\sigma \mid \obs)\,d\sigma,
\end{equation*}
which weights by the posterior $p(\sigma \mid \obs)$ rather than the prior $p(\sigma)$. The two coincide if and only if $p(\sigma \mid \obs) = p(\sigma)$, that is, if the data do not update our beliefs about $\sigma$. We impose this equality as a modelling assumption rather than treating it as an approximation to be checked.

The prior on $\sigma$ represents an external calibration uncertainty as the energy-loss broadening is constrained by auxiliary beamline measurements and expected excitation lifetimes, and $p(\sigma)$ encodes the residual uncertainty in that independent determination. The RIXS spectrum should not override it. Allowing the data to update $\sigma$ could be actively harmful in the presence of model misspecification. Effects absent from the forward model produce spectral features that the simulation cannot reproduce, and a free broadening parameter offers a convenient way to absorb that mismatch: the network can shift $p(\sigma \mid \obs)$ away from the prior to better fit features that have nothing to do with broadening, using $\sigma$ as a pressure-release valve for simulation deficiencies.
Enforcing $p(\sigma \mid \obs) = p(\sigma)$ closes this valve and prevents systematic error from being silently redistributed into a parameter about which we have independent knowledge.

This consideration also dictates how the density estimator is trained.
One might attempt to learn an unconditional model $q_\phi(\param \mid \obs)$ that never sees $\sigma$, but such a network implicitly targets $\int p(\param \mid \obs, \sigma)\,p(\sigma \mid \obs)\,d\sigma$ with the posterior weighting already baked in. 
On simulated training data the spectrum partially identifies $\sigma$, because the simulation obeys the forward model exactly, so the network learns a $p(\sigma \mid \obs) \neq p(\sigma)$. 
Applied to real data, that learned weighting reflects model mismatch rather than genuine information about broadening, and the bias propagates into $p(\param \mid \obs)$. 
Conditioning on $\sigma$ during training and marginalizing over the prior afterwards avoids this entirely as the network is never given the opportunity to learn a data-driven $p(\sigma \mid \obs)$, and the marginalization weight is the externally determined $p(\sigma)$ by construction.

This is the same identifiability issue encountered by \textcite{Lajer2025}, who resolved it by fixing $\sigma$ to a point estimate. 
Integrating over $p(\sigma)$ would generalize their choice from a single value to the full calibration uncertainty while sharing the assumption that the data should not update $\sigma$. 
Empirically the assumption is benign. 
Figure~\ref{fig:nuisance-scan-all} shows that the conditional posteriors shift only slightly as $\sigma$ is varied across its prior range, so even if the data were permitted to update $\sigma$ the effect on $p(\param \mid \obs)$ would be small. 
What the assumption buys is therefore protection against the misspecification pathway above rather than a large numerical correction.

\section{Posterior Mean Parameter Values}
\label{app:posterior-mean}

Table~\ref{tab:posterior-mean} reports the posterior mean $\bar{\param} = \mathbb{E}[\param \mid \obs]$ for both materials. These are the parameter vectors at which the spectrum panels in Figs.~\ref{fig:spectrum_panel} and~\ref{fig:k2nif4-spectra} are evaluated and at which the eigenstate annotations of Appendix~\ref{app:eigenstate} are computed. The nuisance broadening $\sigma$ is excluded as it is marginalized over rather than inferred jointly.

\begin{table}[t]
\caption{Posterior mean Hamiltonian parameters for NiPS$_3$ and K$_2$NiF$_4$.}
\label{tab:posterior-mean}
\centering
\begin{tabular}{@{}lcc@{}}
\toprule
Parameter (eV) & NiPS$_3$ & K$_2$NiF$_4$ \\
\midrule
$F^{(2)}_{dd}$ & 5.333 & 9.013 \\
$F^{(2)}_{dp}$ & 7.901 & 5.055 \\
$F^{(4)}_{dd}$ & 4.572 & 6.237 \\
$G^{(1)}_{dp}$ & 3.612 & 5.377 \\
$G^{(3)}_{dp}$ & 3.214 & 2.547 \\
$\Gamma_c$ & 0.520 & 0.393 \\
$\zeta_c$ & 11.168 & 10.922 \\
$\zeta_{v,i}$ & 0.122 & 0.105 \\
$\zeta_{v,n}$ & 0.110 & 0.097 \\
$10Dq$ & 1.066 & 0.981 \\
\midrule
$\Delta\omega_\mathrm{in}$ & -0.965 & --- \\
$\Delta E$ & --- & -4.553 \\
\bottomrule
\end{tabular}

\end{table}

\section{Software}
\label{app:software}

All of the analysis presented in this work was carried out with open-source scientific software, and we list here the packages on which it depends. The RIXS forward model is evaluated with EDRIXS~\cite{EDRIXS2019}, an open-source exact-diagonalization toolkit for core-hole spectroscopies.

The inference machinery is written in Python~\cite{python3} on top of PyTorch~\cite{pytorch}, with the neural ratio estimators, the conditional flow matching density estimator, and the truncation and active-learning loops implemented directly rather than taken from an existing library. The \texttt{sbi} toolkit~\cite{sbi_toolkit,sbi_reloaded} and \texttt{nflows}~\cite{nflows} nevertheless provided valuable reference implementations against which our own components were cross-checked. Numerical work throughout relies on NumPy~\cite{numpy} and SciPy~\cite{scipy}, with scikit-learn~\cite{scikitlearn} for auxiliary statistical utilities and lmfit~\cite{lmfit} for non-linear least-squares fitting. Simulated and observed spectra are stored and manipulated with h5py~\cite{h5py}, xarray~\cite{xarray}, and pandas~\cite{pandas}, and all figures are produced with Matplotlib~\cite{matplotlib}.

Computational workflows are configured with Hydra~\cite{hydra} and tracked with Weights~\&~Biases~\cite{wandb}. Reproducibility rests on two tools in particular. The full software environment is specified and locked with pixi~\cite{pixi}, so that the exact set of package versions used here can be recreated on another machine. The analysis itself is expressed as a Snakemake workflow~\cite{snakemake}, which encodes every step from simulation through inference to final figure as a declarative dependency graph, so that the results in this paper can be regenerated end to end from the raw inputs by a single invocation rather than by manually rerunning scripts in the correct order. We are grateful to the developers and maintainers of all of these projects, whose work made this study possible.

\bibliography{main}

\begin{thebibliography}{52}%
\makeatletter
\providecommand \@ifxundefined [1]{%
 \@ifx{#1\undefined}
}%
\providecommand \@ifnum [1]{%
 \ifnum #1\expandafter \@firstoftwo
 \else \expandafter \@secondoftwo
 \fi
}%
\providecommand \@ifx [1]{%
 \ifx #1\expandafter \@firstoftwo
 \else \expandafter \@secondoftwo
 \fi
}%
\providecommand \natexlab [1]{#1}%
\providecommand \enquote  [1]{``#1''}%
\providecommand \bibnamefont  [1]{#1}%
\providecommand \bibfnamefont [1]{#1}%
\providecommand \citenamefont [1]{#1}%
\providecommand \href@noop [0]{\@secondoftwo}%
\providecommand \href [0]{\begingroup \@sanitize@url \@href}%
\providecommand \@href[1]{\@@startlink{#1}\@@href}%
\providecommand \@@href[1]{\endgroup#1\@@endlink}%
\providecommand \@sanitize@url [0]{\catcode `\\12\catcode `\$12\catcode `\&12\catcode `\#12\catcode `\^12\catcode `\_12\catcode `\%12\relax}%
\providecommand \@@startlink[1]{}%
\providecommand \@@endlink[0]{}%
\providecommand \url  [0]{\begingroup\@sanitize@url \@url }%
\providecommand \@url [1]{\endgroup\@href {#1}{\urlprefix }}%
\providecommand \urlprefix  [0]{URL }%
\providecommand \Eprint [0]{\href }%
\providecommand \doibase [0]{https://doi.org/}%
\providecommand \selectlanguage [0]{\@gobble}%
\providecommand \bibinfo  [0]{\@secondoftwo}%
\providecommand \bibfield  [0]{\@secondoftwo}%
\providecommand \translation [1]{[#1]}%
\providecommand \BibitemOpen [0]{}%
\providecommand \bibitemStop [0]{}%
\providecommand \bibitemNoStop [0]{.\EOS\space}%
\providecommand \EOS [0]{\spacefactor3000\relax}%
\providecommand \BibitemShut  [1]{\csname bibitem#1\endcsname}%
\let\auto@bib@innerbib\@empty
\bibitem [{\citenamefont {Ament}\ \emph {et~al.}(2011)\citenamefont {Ament}, \citenamefont {van Veenendaal}, \citenamefont {Devereaux}, \citenamefont {Hill},\ and\ \citenamefont {van~den Brink}}]{Ament2011}%
  \BibitemOpen
  \bibfield  {author} {\bibinfo {author} {\bibfnamefont {L.~J.~P.}\ \bibnamefont {Ament}}, \bibinfo {author} {\bibfnamefont {M.}~\bibnamefont {van Veenendaal}}, \bibinfo {author} {\bibfnamefont {T.~P.}\ \bibnamefont {Devereaux}}, \bibinfo {author} {\bibfnamefont {J.~P.}\ \bibnamefont {Hill}},\ and\ \bibinfo {author} {\bibfnamefont {J.}~\bibnamefont {van~den Brink}},\ }\bibfield  {title} {\bibinfo {title} {Resonant inelastic x-ray scattering studies of elementary excitations},\ }\href {https://doi.org/10.1103/RevModPhys.83.705} {\bibfield  {journal} {\bibinfo  {journal} {Reviews of Modern Physics}\ }\textbf {\bibinfo {volume} {83}},\ \bibinfo {pages} {705} (\bibinfo {year} {2011})}\BibitemShut {NoStop}%
\bibitem [{\citenamefont {Kotani}\ and\ \citenamefont {Shin}(2001)}]{Kotani2001}%
  \BibitemOpen
  \bibfield  {author} {\bibinfo {author} {\bibfnamefont {A.}~\bibnamefont {Kotani}}\ and\ \bibinfo {author} {\bibfnamefont {S.}~\bibnamefont {Shin}},\ }\bibfield  {title} {\bibinfo {title} {Resonant inelastic x-ray scattering spectra for electrons in solids},\ }\href {https://doi.org/10.1103/RevModPhys.73.203} {\bibfield  {journal} {\bibinfo  {journal} {Reviews of Modern Physics}\ }\textbf {\bibinfo {volume} {73}},\ \bibinfo {pages} {203} (\bibinfo {year} {2001})}\BibitemShut {NoStop}%
\bibitem [{\citenamefont {Dean}(2015)}]{Dean2015}%
  \BibitemOpen
  \bibfield  {author} {\bibinfo {author} {\bibfnamefont {M.~P.~M.}\ \bibnamefont {Dean}},\ }\bibfield  {title} {\bibinfo {title} {Insights into the high temperature superconducting cuprates from resonant inelastic x-ray scattering},\ }\href {https://doi.org/10.1016/j.jmmm.2014.03.057} {\bibfield  {journal} {\bibinfo  {journal} {Journal of Magnetism and Magnetic Materials}\ }\textbf {\bibinfo {volume} {376}},\ \bibinfo {pages} {3} (\bibinfo {year} {2015})}\BibitemShut {NoStop}%
\bibitem [{\citenamefont {de~Groot}\ \emph {et~al.}(2024)\citenamefont {de~Groot}, \citenamefont {Haverkort}, \citenamefont {Elnaggar}, \citenamefont {Juhin}, \citenamefont {Zhou},\ and\ \citenamefont {Glatzel}}]{deGroot2024}%
  \BibitemOpen
  \bibfield  {author} {\bibinfo {author} {\bibfnamefont {F.~M.~F.}\ \bibnamefont {de~Groot}}, \bibinfo {author} {\bibfnamefont {M.~W.}\ \bibnamefont {Haverkort}}, \bibinfo {author} {\bibfnamefont {H.}~\bibnamefont {Elnaggar}}, \bibinfo {author} {\bibfnamefont {A.}~\bibnamefont {Juhin}}, \bibinfo {author} {\bibfnamefont {K.-J.}\ \bibnamefont {Zhou}},\ and\ \bibinfo {author} {\bibfnamefont {P.}~\bibnamefont {Glatzel}},\ }\bibfield  {title} {\bibinfo {title} {Resonant inelastic {X}-ray scattering},\ }\href {https://doi.org/10.1038/s43586-024-00322-6} {\bibfield  {journal} {\bibinfo  {journal} {Nature Reviews Methods Primers}\ }\textbf {\bibinfo {volume} {4}},\ \bibinfo {pages} {45} (\bibinfo {year} {2024})}\BibitemShut {NoStop}%
\bibitem [{\citenamefont {Mitrano}\ \emph {et~al.}(2024)\citenamefont {Mitrano}, \citenamefont {Johnston}, \citenamefont {Kim},\ and\ \citenamefont {Dean}}]{Mitrano2024exploring}%
  \BibitemOpen
  \bibfield  {author} {\bibinfo {author} {\bibfnamefont {M.}~\bibnamefont {Mitrano}}, \bibinfo {author} {\bibfnamefont {S.}~\bibnamefont {Johnston}}, \bibinfo {author} {\bibfnamefont {Y.-J.}\ \bibnamefont {Kim}},\ and\ \bibinfo {author} {\bibfnamefont {M.~P.~M.}\ \bibnamefont {Dean}},\ }\bibfield  {title} {\bibinfo {title} {{Exploring Quantum Materials with Resonant Inelastic X-Ray Scattering}},\ }\href {https://doi.org/10.1103/PhysRevX.14.040501} {\bibfield  {journal} {\bibinfo  {journal} {Phys. Rev. X}\ }\textbf {\bibinfo {volume} {14}},\ \bibinfo {pages} {040501} (\bibinfo {year} {2024})}\BibitemShut {NoStop}%
\bibitem [{\citenamefont {de~Groot}\ and\ \citenamefont {Kotani}(2008)}]{deGroot2008}%
  \BibitemOpen
  \bibfield  {author} {\bibinfo {author} {\bibfnamefont {F.}~\bibnamefont {de~Groot}}\ and\ \bibinfo {author} {\bibfnamefont {A.}~\bibnamefont {Kotani}},\ }\href {https://doi.org/10.1201/9781420008425} {\emph {\bibinfo {title} {Core Level Spectroscopy of Solids}}},\ Advances in Condensed Matter Science\ (\bibinfo  {publisher} {CRC Press},\ \bibinfo {address} {Boca Raton, FL},\ \bibinfo {year} {2008})\BibitemShut {NoStop}%
\bibitem [{\citenamefont {Haverkort}\ \emph {et~al.}(2012)\citenamefont {Haverkort}, \citenamefont {Zwierzycki},\ and\ \citenamefont {Andersen}}]{Haverkort2012}%
  \BibitemOpen
  \bibfield  {author} {\bibinfo {author} {\bibfnamefont {M.~W.}\ \bibnamefont {Haverkort}}, \bibinfo {author} {\bibfnamefont {M.}~\bibnamefont {Zwierzycki}},\ and\ \bibinfo {author} {\bibfnamefont {O.~K.}\ \bibnamefont {Andersen}},\ }\bibfield  {title} {\bibinfo {title} {Multiplet ligand-field theory using {Wannier} orbitals},\ }\href {https://doi.org/10.1103/PhysRevB.85.165113} {\bibfield  {journal} {\bibinfo  {journal} {Physical Review B}\ }\textbf {\bibinfo {volume} {85}},\ \bibinfo {pages} {165113} (\bibinfo {year} {2012})}\BibitemShut {NoStop}%
\bibitem [{\citenamefont {Wang}\ \emph {et~al.}(2019)\citenamefont {Wang}, \citenamefont {Fabbris}, \citenamefont {Dean},\ and\ \citenamefont {Kotliar}}]{EDRIXS2019}%
  \BibitemOpen
  \bibfield  {author} {\bibinfo {author} {\bibfnamefont {Y.~L.}\ \bibnamefont {Wang}}, \bibinfo {author} {\bibfnamefont {G.}~\bibnamefont {Fabbris}}, \bibinfo {author} {\bibfnamefont {M.~P.~M.}\ \bibnamefont {Dean}},\ and\ \bibinfo {author} {\bibfnamefont {G.}~\bibnamefont {Kotliar}},\ }\bibfield  {title} {\bibinfo {title} {{EDRIXS}: An open source toolkit for simulating spectra of resonant inelastic x-ray scattering},\ }\href {https://doi.org/10.1016/j.cpc.2019.04.018} {\bibfield  {journal} {\bibinfo  {journal} {Computer Physics Communications}\ }\textbf {\bibinfo {volume} {243}},\ \bibinfo {pages} {151} (\bibinfo {year} {2019})}\BibitemShut {NoStop}%
\bibitem [{\citenamefont {Lajer}\ \emph {et~al.}(2025)\citenamefont {Lajer}, \citenamefont {Dai}, \citenamefont {Barros}, \citenamefont {Carbone}, \citenamefont {Johnston},\ and\ \citenamefont {Dean}}]{Lajer2025}%
  \BibitemOpen
  \bibfield  {author} {\bibinfo {author} {\bibfnamefont {M.~K.}\ \bibnamefont {Lajer}}, \bibinfo {author} {\bibfnamefont {X.}~\bibnamefont {Dai}}, \bibinfo {author} {\bibfnamefont {K.}~\bibnamefont {Barros}}, \bibinfo {author} {\bibfnamefont {M.~R.}\ \bibnamefont {Carbone}}, \bibinfo {author} {\bibfnamefont {S.}~\bibnamefont {Johnston}},\ and\ \bibinfo {author} {\bibfnamefont {M.~P.~M.}\ \bibnamefont {Dean}},\ }\bibfield  {title} {\bibinfo {title} {Hamiltonian parameter inference from resonant inelastic x-ray scattering with active learning},\ }\href {https://doi.org/10.1103/tnqm-ttj3} {\bibfield  {journal} {\bibinfo  {journal} {Phys. Rev. B}\ }\textbf {\bibinfo {volume} {112}},\ \bibinfo {pages} {155167} (\bibinfo {year} {2025})}\BibitemShut {NoStop}%
\bibitem [{\citenamefont {Miller}\ \emph {et~al.}(2021)\citenamefont {Miller}, \citenamefont {Cole}, \citenamefont {Forr{\'e}}, \citenamefont {Louppe},\ and\ \citenamefont {Weniger}}]{miller2021truncated}%
  \BibitemOpen
  \bibfield  {author} {\bibinfo {author} {\bibfnamefont {B.~K.}\ \bibnamefont {Miller}}, \bibinfo {author} {\bibfnamefont {A.}~\bibnamefont {Cole}}, \bibinfo {author} {\bibfnamefont {P.}~\bibnamefont {Forr{\'e}}}, \bibinfo {author} {\bibfnamefont {G.}~\bibnamefont {Louppe}},\ and\ \bibinfo {author} {\bibfnamefont {C.}~\bibnamefont {Weniger}},\ }\bibfield  {title} {\bibinfo {title} {Truncated marginal neural ratio estimation},\ }\href@noop {} {\bibfield  {journal} {\bibinfo  {journal} {Advances in neural information processing systems}\ }\textbf {\bibinfo {volume} {34}},\ \bibinfo {pages} {129} (\bibinfo {year} {2021})}\BibitemShut {NoStop}%
\bibitem [{\citenamefont {Cranmer}\ \emph {et~al.}(2020)\citenamefont {Cranmer}, \citenamefont {Brehmer},\ and\ \citenamefont {Louppe}}]{Cranmer2020}%
  \BibitemOpen
  \bibfield  {author} {\bibinfo {author} {\bibfnamefont {K.}~\bibnamefont {Cranmer}}, \bibinfo {author} {\bibfnamefont {J.}~\bibnamefont {Brehmer}},\ and\ \bibinfo {author} {\bibfnamefont {G.}~\bibnamefont {Louppe}},\ }\bibfield  {title} {\bibinfo {title} {The frontier of simulation-based inference},\ }\href {https://doi.org/10.1073/pnas.1912789117} {\bibfield  {journal} {\bibinfo  {journal} {Proceedings of the National Academy of Sciences}\ }\textbf {\bibinfo {volume} {117}},\ \bibinfo {pages} {30055} (\bibinfo {year} {2020})}\BibitemShut {NoStop}%
\bibitem [{\citenamefont {Hermans}\ \emph {et~al.}(2020)\citenamefont {Hermans}, \citenamefont {Begy},\ and\ \citenamefont {Louppe}}]{Hermans2020}%
  \BibitemOpen
  \bibfield  {author} {\bibinfo {author} {\bibfnamefont {J.}~\bibnamefont {Hermans}}, \bibinfo {author} {\bibfnamefont {V.}~\bibnamefont {Begy}},\ and\ \bibinfo {author} {\bibfnamefont {G.}~\bibnamefont {Louppe}},\ }\bibfield  {title} {\bibinfo {title} {Likelihood-free {MCMC} with amortized approximate ratio estimators},\ }in\ \href@noop {} {\emph {\bibinfo {booktitle} {International Conference on Machine Learning}}}\ (\bibinfo {organization} {PMLR},\ \bibinfo {year} {2020})\ pp.\ \bibinfo {pages} {4239--4248}\BibitemShut {NoStop}%
\bibitem [{\citenamefont {Papamakarios}\ and\ \citenamefont {Murray}(2016)}]{Papamakarios2016}%
  \BibitemOpen
  \bibfield  {author} {\bibinfo {author} {\bibfnamefont {G.}~\bibnamefont {Papamakarios}}\ and\ \bibinfo {author} {\bibfnamefont {I.}~\bibnamefont {Murray}},\ }\bibfield  {title} {\bibinfo {title} {Fast $\varepsilon$-free inference of simulation models with {B}ayesian conditional density estimation},\ }in\ \href@noop {} {\emph {\bibinfo {booktitle} {Advances in Neural Information Processing Systems}}},\ Vol.~\bibinfo {volume} {29}\ (\bibinfo {year} {2016})\BibitemShut {NoStop}%
\bibitem [{\citenamefont {Lipman}\ \emph {et~al.}(2023)\citenamefont {Lipman}, \citenamefont {Chen}, \citenamefont {Ben-Hamu}, \citenamefont {Nickel},\ and\ \citenamefont {Le}}]{Lipman2023}%
  \BibitemOpen
  \bibfield  {author} {\bibinfo {author} {\bibfnamefont {Y.}~\bibnamefont {Lipman}}, \bibinfo {author} {\bibfnamefont {R.~T.~Q.}\ \bibnamefont {Chen}}, \bibinfo {author} {\bibfnamefont {H.}~\bibnamefont {Ben-Hamu}}, \bibinfo {author} {\bibfnamefont {M.}~\bibnamefont {Nickel}},\ and\ \bibinfo {author} {\bibfnamefont {M.}~\bibnamefont {Le}},\ }\bibfield  {title} {\bibinfo {title} {Flow matching for generative modeling},\ }in\ \href@noop {} {\emph {\bibinfo {booktitle} {International Conference on Learning Representations}}}\ (\bibinfo {year} {2023})\BibitemShut {NoStop}%
\bibitem [{\citenamefont {Albergo}\ \emph {et~al.}(2023)\citenamefont {Albergo}, \citenamefont {Boffi},\ and\ \citenamefont {Vanden-Eijnden}}]{Albergo2023}%
  \BibitemOpen
  \bibfield  {author} {\bibinfo {author} {\bibfnamefont {M.~S.}\ \bibnamefont {Albergo}}, \bibinfo {author} {\bibfnamefont {N.~M.}\ \bibnamefont {Boffi}},\ and\ \bibinfo {author} {\bibfnamefont {E.}~\bibnamefont {Vanden-Eijnden}},\ }\bibfield  {title} {\bibinfo {title} {Stochastic interpolants: A unifying framework for flows and diffusions},\ }in\ \href@noop {} {\emph {\bibinfo {booktitle} {International Conference on Machine Learning}}}\ (\bibinfo {organization} {PMLR},\ \bibinfo {year} {2023})\BibitemShut {NoStop}%
\bibitem [{\citenamefont {Cowan}(1981)}]{Cowan1981}%
  \BibitemOpen
  \bibfield  {author} {\bibinfo {author} {\bibfnamefont {R.~D.}\ \bibnamefont {Cowan}},\ }\href@noop {} {\emph {\bibinfo {title} {The Theory of Atomic Structure and Spectra}}}\ (\bibinfo  {publisher} {University of California Press},\ \bibinfo {address} {Berkeley},\ \bibinfo {year} {1981})\BibitemShut {NoStop}%
\bibitem [{\citenamefont {Shannon}(1976)}]{shannon1976revised}%
  \BibitemOpen
  \bibfield  {author} {\bibinfo {author} {\bibfnamefont {R.~D.}\ \bibnamefont {Shannon}},\ }\bibfield  {title} {\bibinfo {title} {Revised effective ionic radii and systematic studies of interatomic distances in halides and chalcogenides},\ }\href@noop {} {\bibfield  {journal} {\bibinfo  {journal} {Foundations of Crystallography}\ }\textbf {\bibinfo {volume} {32}},\ \bibinfo {pages} {751} (\bibinfo {year} {1976})}\BibitemShut {NoStop}%
\bibitem [{\citenamefont {Van~der Laan}\ \emph {et~al.}(1988)\citenamefont {Van~der Laan}, \citenamefont {Thole}, \citenamefont {Sawatzky},\ and\ \citenamefont {Verdaguer}}]{van1988multiplet}%
  \BibitemOpen
  \bibfield  {author} {\bibinfo {author} {\bibfnamefont {G.}~\bibnamefont {Van~der Laan}}, \bibinfo {author} {\bibfnamefont {B.}~\bibnamefont {Thole}}, \bibinfo {author} {\bibfnamefont {G.}~\bibnamefont {Sawatzky}},\ and\ \bibinfo {author} {\bibfnamefont {M.}~\bibnamefont {Verdaguer}},\ }\bibfield  {title} {\bibinfo {title} {Multiplet structure in the l 2, 3 x-ray-absorption spectra: A fingerprint for high-and low-spin ni 2+ compounds},\ }\href@noop {} {\bibfield  {journal} {\bibinfo  {journal} {Physical Review B}\ }\textbf {\bibinfo {volume} {37}},\ \bibinfo {pages} {6587} (\bibinfo {year} {1988})}\BibitemShut {NoStop}%
\bibitem [{\citenamefont {He}\ \emph {et~al.}(2016)\citenamefont {He}, \citenamefont {Zhang}, \citenamefont {Ren},\ and\ \citenamefont {Sun}}]{He2016}%
  \BibitemOpen
  \bibfield  {author} {\bibinfo {author} {\bibfnamefont {K.}~\bibnamefont {He}}, \bibinfo {author} {\bibfnamefont {X.}~\bibnamefont {Zhang}}, \bibinfo {author} {\bibfnamefont {S.}~\bibnamefont {Ren}},\ and\ \bibinfo {author} {\bibfnamefont {J.}~\bibnamefont {Sun}},\ }\bibfield  {title} {\bibinfo {title} {Deep residual learning for image recognition},\ }in\ \href {https://doi.org/10.1109/CVPR.2016.90} {\emph {\bibinfo {booktitle} {Proceedings of the IEEE Conference on Computer Vision and Pattern Recognition (CVPR)}}}\ (\bibinfo {year} {2016})\ pp.\ \bibinfo {pages} {770--778}\BibitemShut {NoStop}%
\bibitem [{\citenamefont {Dosovitskiy}\ \emph {et~al.}(2021)\citenamefont {Dosovitskiy}, \citenamefont {Beyer}, \citenamefont {Kolesnikov}, \citenamefont {Weissenborn}, \citenamefont {Zhai}, \citenamefont {Unterthiner}, \citenamefont {Dehghani}, \citenamefont {Minderer}, \citenamefont {Heigold}, \citenamefont {Gelly}, \citenamefont {Uszkoreit},\ and\ \citenamefont {Houlsby}}]{Dosovitskiy2021}%
  \BibitemOpen
  \bibfield  {author} {\bibinfo {author} {\bibfnamefont {A.}~\bibnamefont {Dosovitskiy}}, \bibinfo {author} {\bibfnamefont {L.}~\bibnamefont {Beyer}}, \bibinfo {author} {\bibfnamefont {A.}~\bibnamefont {Kolesnikov}}, \bibinfo {author} {\bibfnamefont {D.}~\bibnamefont {Weissenborn}}, \bibinfo {author} {\bibfnamefont {X.}~\bibnamefont {Zhai}}, \bibinfo {author} {\bibfnamefont {T.}~\bibnamefont {Unterthiner}}, \bibinfo {author} {\bibfnamefont {M.}~\bibnamefont {Dehghani}}, \bibinfo {author} {\bibfnamefont {M.}~\bibnamefont {Minderer}}, \bibinfo {author} {\bibfnamefont {G.}~\bibnamefont {Heigold}}, \bibinfo {author} {\bibfnamefont {S.}~\bibnamefont {Gelly}}, \bibinfo {author} {\bibfnamefont {J.}~\bibnamefont {Uszkoreit}},\ and\ \bibinfo {author} {\bibfnamefont {N.}~\bibnamefont {Houlsby}},\ }\bibfield  {title} {\bibinfo {title} {An image is worth 16x16 words: Transformers for image recognition at scale},\ }in\ \href@noop {} {\emph {\bibinfo {booktitle} {International Conference on Learning Representations (ICLR)}}}\ (\bibinfo {year} {2021})\BibitemShut {NoStop}%
\bibitem [{\citenamefont {Vaswani}\ \emph {et~al.}(2017)\citenamefont {Vaswani}, \citenamefont {Shazeer}, \citenamefont {Parmar}, \citenamefont {Uszkoreit}, \citenamefont {Jones}, \citenamefont {Gomez}, \citenamefont {Kaiser},\ and\ \citenamefont {Polosukhin}}]{Vaswani2017}%
  \BibitemOpen
  \bibfield  {author} {\bibinfo {author} {\bibfnamefont {A.}~\bibnamefont {Vaswani}}, \bibinfo {author} {\bibfnamefont {N.}~\bibnamefont {Shazeer}}, \bibinfo {author} {\bibfnamefont {N.}~\bibnamefont {Parmar}}, \bibinfo {author} {\bibfnamefont {J.}~\bibnamefont {Uszkoreit}}, \bibinfo {author} {\bibfnamefont {L.}~\bibnamefont {Jones}}, \bibinfo {author} {\bibfnamefont {A.~N.}\ \bibnamefont {Gomez}}, \bibinfo {author} {\bibfnamefont {{\L}.}~\bibnamefont {Kaiser}},\ and\ \bibinfo {author} {\bibfnamefont {I.}~\bibnamefont {Polosukhin}},\ }\bibfield  {title} {\bibinfo {title} {Attention is all you need},\ }in\ \href@noop {} {\emph {\bibinfo {booktitle} {Advances in Neural Information Processing Systems}}},\ Vol.~\bibinfo {volume} {30}\ (\bibinfo {year} {2017})\BibitemShut {NoStop}%
\bibitem [{\citenamefont {Devlin}\ \emph {et~al.}(2019)\citenamefont {Devlin}, \citenamefont {Chang}, \citenamefont {Lee},\ and\ \citenamefont {Toutanova}}]{Devlin2019}%
  \BibitemOpen
  \bibfield  {author} {\bibinfo {author} {\bibfnamefont {J.}~\bibnamefont {Devlin}}, \bibinfo {author} {\bibfnamefont {M.-W.}\ \bibnamefont {Chang}}, \bibinfo {author} {\bibfnamefont {K.}~\bibnamefont {Lee}},\ and\ \bibinfo {author} {\bibfnamefont {K.}~\bibnamefont {Toutanova}},\ }\bibfield  {title} {\bibinfo {title} {{BERT}: Pre-training of deep bidirectional transformers for language understanding},\ }in\ \href {https://doi.org/10.18653/v1/N19-1423} {\emph {\bibinfo {booktitle} {Proceedings of the 2019 Conference of the North {A}merican Chapter of the Association for Computational Linguistics: Human Language Technologies}}}\ (\bibinfo {year} {2019})\ pp.\ \bibinfo {pages} {4171--4186}\BibitemShut {NoStop}%
\bibitem [{\citenamefont {Lakshminarayanan}\ \emph {et~al.}(2017)\citenamefont {Lakshminarayanan}, \citenamefont {Pritzel},\ and\ \citenamefont {Blundell}}]{Lakshminarayanan2017}%
  \BibitemOpen
  \bibfield  {author} {\bibinfo {author} {\bibfnamefont {B.}~\bibnamefont {Lakshminarayanan}}, \bibinfo {author} {\bibfnamefont {A.}~\bibnamefont {Pritzel}},\ and\ \bibinfo {author} {\bibfnamefont {C.}~\bibnamefont {Blundell}},\ }\bibfield  {title} {\bibinfo {title} {Simple and scalable predictive uncertainty estimation using deep ensembles},\ }\href@noop {} {\bibfield  {journal} {\bibinfo  {journal} {Advances in Neural Information Processing Systems}\ }\textbf {\bibinfo {volume} {30}} (\bibinfo {year} {2017})}\BibitemShut {NoStop}%
\bibitem [{\citenamefont {Hermans}\ \emph {et~al.}(2022)\citenamefont {Hermans}, \citenamefont {Delaunoy}, \citenamefont {Rozet}, \citenamefont {Wehenkel}, \citenamefont {Begy},\ and\ \citenamefont {Louppe}}]{Hermans2022crisis}%
  \BibitemOpen
  \bibfield  {author} {\bibinfo {author} {\bibfnamefont {J.}~\bibnamefont {Hermans}}, \bibinfo {author} {\bibfnamefont {A.}~\bibnamefont {Delaunoy}}, \bibinfo {author} {\bibfnamefont {F.}~\bibnamefont {Rozet}}, \bibinfo {author} {\bibfnamefont {A.}~\bibnamefont {Wehenkel}}, \bibinfo {author} {\bibfnamefont {V.}~\bibnamefont {Begy}},\ and\ \bibinfo {author} {\bibfnamefont {G.}~\bibnamefont {Louppe}},\ }\bibfield  {title} {\bibinfo {title} {A crisis in simulation-based inference? {B}eware, your posterior approximations can be unfaithful},\ }in\ \href@noop {} {\emph {\bibinfo {booktitle} {Transactions on Machine Learning Research}}}\ (\bibinfo {year} {2022})\BibitemShut {NoStop}%
\bibitem [{\citenamefont {Wildes}\ \emph {et~al.}(2015)\citenamefont {Wildes}, \citenamefont {Simonet}, \citenamefont {Ressouche}, \citenamefont {McIntyre}, \citenamefont {Avdeev}, \citenamefont {Suard}, \citenamefont {Kimber}, \citenamefont {Lan{\c{c}}on}, \citenamefont {Pepe}, \citenamefont {Moubaraki},\ and\ \citenamefont {Hicks}}]{Wildes2015}%
  \BibitemOpen
  \bibfield  {author} {\bibinfo {author} {\bibfnamefont {A.~R.}\ \bibnamefont {Wildes}}, \bibinfo {author} {\bibfnamefont {V.}~\bibnamefont {Simonet}}, \bibinfo {author} {\bibfnamefont {E.}~\bibnamefont {Ressouche}}, \bibinfo {author} {\bibfnamefont {G.~J.}\ \bibnamefont {McIntyre}}, \bibinfo {author} {\bibfnamefont {M.}~\bibnamefont {Avdeev}}, \bibinfo {author} {\bibfnamefont {E.}~\bibnamefont {Suard}}, \bibinfo {author} {\bibfnamefont {S.~A.~J.}\ \bibnamefont {Kimber}}, \bibinfo {author} {\bibfnamefont {D.}~\bibnamefont {Lan{\c{c}}on}}, \bibinfo {author} {\bibfnamefont {G.}~\bibnamefont {Pepe}}, \bibinfo {author} {\bibfnamefont {B.}~\bibnamefont {Moubaraki}},\ and\ \bibinfo {author} {\bibfnamefont {T.~J.}\ \bibnamefont {Hicks}},\ }\bibfield  {title} {\bibinfo {title} {Magnetic structure of the quasi-two-dimensional antiferromagnet {NiPS}$_3$},\ }\href {https://doi.org/10.1103/PhysRevB.92.224408} {\bibfield  {journal} {\bibinfo  {journal} {Physical Review B}\ }\textbf {\bibinfo {volume} {92}},\ \bibinfo {pages} {224408} (\bibinfo {year} {2015})}\BibitemShut {NoStop}%
\bibitem [{\citenamefont {Kang}\ \emph {et~al.}(2020)\citenamefont {Kang}, \citenamefont {Kim}, \citenamefont {Kim}, \citenamefont {Kim}, \citenamefont {Sim}, \citenamefont {Lee}, \citenamefont {Lee}, \citenamefont {Park}, \citenamefont {Yun}, \citenamefont {Kim}, \citenamefont {Nag}, \citenamefont {Walters}, \citenamefont {Garcia-Fernandez}, \citenamefont {Li}, \citenamefont {Chapon}, \citenamefont {Zhou}, \citenamefont {Son}, \citenamefont {Kim}, \citenamefont {Cheong},\ and\ \citenamefont {Park}}]{Kang2020}%
  \BibitemOpen
  \bibfield  {author} {\bibinfo {author} {\bibfnamefont {S.}~\bibnamefont {Kang}}, \bibinfo {author} {\bibfnamefont {K.}~\bibnamefont {Kim}}, \bibinfo {author} {\bibfnamefont {B.~H.}\ \bibnamefont {Kim}}, \bibinfo {author} {\bibfnamefont {J.}~\bibnamefont {Kim}}, \bibinfo {author} {\bibfnamefont {K.~I.}\ \bibnamefont {Sim}}, \bibinfo {author} {\bibfnamefont {J.-U.}\ \bibnamefont {Lee}}, \bibinfo {author} {\bibfnamefont {S.}~\bibnamefont {Lee}}, \bibinfo {author} {\bibfnamefont {K.}~\bibnamefont {Park}}, \bibinfo {author} {\bibfnamefont {S.}~\bibnamefont {Yun}}, \bibinfo {author} {\bibfnamefont {T.}~\bibnamefont {Kim}}, \bibinfo {author} {\bibfnamefont {A.}~\bibnamefont {Nag}}, \bibinfo {author} {\bibfnamefont {A.}~\bibnamefont {Walters}}, \bibinfo {author} {\bibfnamefont {M.}~\bibnamefont {Garcia-Fernandez}}, \bibinfo {author} {\bibfnamefont {J.}~\bibnamefont {Li}}, \bibinfo {author} {\bibfnamefont {L.}~\bibnamefont {Chapon}}, \bibinfo {author} {\bibfnamefont {K.-J.}\ \bibnamefont {Zhou}}, \bibinfo {author} {\bibfnamefont {Y.-W.}\ \bibnamefont {Son}}, \bibinfo {author} {\bibfnamefont {J.~H.}\ \bibnamefont {Kim}}, \bibinfo {author} {\bibfnamefont {H.}~\bibnamefont {Cheong}},\ and\ \bibinfo {author} {\bibfnamefont {J.-G.}\ \bibnamefont {Park}},\ }\bibfield  {title} {\bibinfo {title} {Coherent many-body exciton in van der waals antiferromagnet {NiPS$_3$}},\ }\href {https://doi.org/10.1038/s41586-020-2520-5} {\bibfield  {journal} {\bibinfo  {journal} {Nature}\ }\textbf {\bibinfo {volume} {583}},\ \bibinfo {pages} {785} (\bibinfo {year} {2020})}\BibitemShut {NoStop}%
\bibitem [{\citenamefont {He}\ \emph {et~al.}(2024)\citenamefont {He}, \citenamefont {Shen}, \citenamefont {Wohlfeld}, \citenamefont {Sears}, \citenamefont {Li}, \citenamefont {Pelliciari}, \citenamefont {Walicki}, \citenamefont {Johnston}, \citenamefont {Baldini}, \citenamefont {Bisogni}, \citenamefont {Mitrano},\ and\ \citenamefont {Dean}}]{He2024}%
  \BibitemOpen
  \bibfield  {author} {\bibinfo {author} {\bibfnamefont {W.}~\bibnamefont {He}}, \bibinfo {author} {\bibfnamefont {Y.}~\bibnamefont {Shen}}, \bibinfo {author} {\bibfnamefont {K.}~\bibnamefont {Wohlfeld}}, \bibinfo {author} {\bibfnamefont {J.}~\bibnamefont {Sears}}, \bibinfo {author} {\bibfnamefont {J.}~\bibnamefont {Li}}, \bibinfo {author} {\bibfnamefont {J.}~\bibnamefont {Pelliciari}}, \bibinfo {author} {\bibfnamefont {M.}~\bibnamefont {Walicki}}, \bibinfo {author} {\bibfnamefont {S.}~\bibnamefont {Johnston}}, \bibinfo {author} {\bibfnamefont {E.}~\bibnamefont {Baldini}}, \bibinfo {author} {\bibfnamefont {V.}~\bibnamefont {Bisogni}}, \bibinfo {author} {\bibfnamefont {M.}~\bibnamefont {Mitrano}},\ and\ \bibinfo {author} {\bibfnamefont {M.~P.~M.}\ \bibnamefont {Dean}},\ }\bibfield  {title} {\bibinfo {title} {Magnetically propagating hund's exciton in van der waals antiferromagnet nips3},\ }\href {https://doi.org/10.1038/s41467-024-47852-x} {\bibfield  {journal} {\bibinfo  {journal} {Nature Communications}\ }\textbf {\bibinfo {volume} {15}},\ \bibinfo {pages} {3496} (\bibinfo {year} {2024})}\BibitemShut {NoStop}%
\bibitem [{\citenamefont {Cartis}\ \emph {et~al.}(2018)\citenamefont {Cartis}, \citenamefont {Fiala}, \citenamefont {Marteau},\ and\ \citenamefont {Roberts}}]{cartis2018improvingflexibilityrobustnessmodelbased}%
  \BibitemOpen
  \bibfield  {author} {\bibinfo {author} {\bibfnamefont {C.}~\bibnamefont {Cartis}}, \bibinfo {author} {\bibfnamefont {J.}~\bibnamefont {Fiala}}, \bibinfo {author} {\bibfnamefont {B.}~\bibnamefont {Marteau}},\ and\ \bibinfo {author} {\bibfnamefont {L.}~\bibnamefont {Roberts}},\ }\href {https://arxiv.org/abs/1804.00154} {\bibinfo {title} {Improving the flexibility and robustness of model-based derivative-free optimization solvers}} (\bibinfo {year} {2018}),\ \Eprint {https://arxiv.org/abs/1804.00154} {arXiv:1804.00154 [math.OC]} \BibitemShut {NoStop}%
\bibitem [{\citenamefont {Lemos}\ \emph {et~al.}(2023)\citenamefont {Lemos}, \citenamefont {Coogan}, \citenamefont {Hezaveh},\ and\ \citenamefont {Perreault-Levasseur}}]{Lemos2023}%
  \BibitemOpen
  \bibfield  {author} {\bibinfo {author} {\bibfnamefont {P.}~\bibnamefont {Lemos}}, \bibinfo {author} {\bibfnamefont {A.}~\bibnamefont {Coogan}}, \bibinfo {author} {\bibfnamefont {Y.}~\bibnamefont {Hezaveh}},\ and\ \bibinfo {author} {\bibfnamefont {L.}~\bibnamefont {Perreault-Levasseur}},\ }\href@noop {} {\bibinfo {title} {{Sampling-Based Accuracy Testing of Posterior Estimators for General Inference}}} (\bibinfo {year} {2023}),\ \Eprint {https://arxiv.org/abs/2302.03026} {arXiv:2302.03026 [stat.ML]} \BibitemShut {NoStop}%
\bibitem [{\citenamefont {Wanklyn}(1975)}]{Wanklyn1975flux}%
  \BibitemOpen
  \bibfield  {author} {\bibinfo {author} {\bibfnamefont {B.~M.}\ \bibnamefont {Wanklyn}},\ }\bibfield  {title} {\bibinfo {title} {Flux growth of crystals of some transition metal fluorides},\ }\href {https://doi.org/10.1007/BF01031848} {\bibfield  {journal} {\bibinfo  {journal} {Journal of Materials Science}\ }\textbf {\bibinfo {volume} {10}},\ \bibinfo {pages} {1487} (\bibinfo {year} {1975})}\BibitemShut {NoStop}%
\bibitem [{\citenamefont {Klein}\ \emph {et~al.}(2026{\natexlab{a}})\citenamefont {Klein}, \citenamefont {Linker}, \citenamefont {Conreux}, \citenamefont {Ratner}, \citenamefont {Mehta}, \citenamefont {Tachibana}, \citenamefont {Li}, \citenamefont {Pelliciari}, \citenamefont {Bisogni}, \citenamefont {He}, \citenamefont {Luo}, \citenamefont {Dean}, \citenamefont {Lajer}, \citenamefont {Kagan}, \citenamefont {Turner}, \citenamefont {Cheng},\ and\ \citenamefont {Gasiorowski}}]{klein_2026_21907963}%
  \BibitemOpen
  \bibfield  {author} {\bibinfo {author} {\bibfnamefont {S.}~\bibnamefont {Klein}}, \bibinfo {author} {\bibfnamefont {T.~M.}\ \bibnamefont {Linker}}, \bibinfo {author} {\bibfnamefont {L.}~\bibnamefont {Conreux}}, \bibinfo {author} {\bibfnamefont {D.}~\bibnamefont {Ratner}}, \bibinfo {author} {\bibfnamefont {A.}~\bibnamefont {Mehta}}, \bibinfo {author} {\bibfnamefont {M.}~\bibnamefont {Tachibana}}, \bibinfo {author} {\bibfnamefont {J.}~\bibnamefont {Li}}, \bibinfo {author} {\bibfnamefont {J.}~\bibnamefont {Pelliciari}}, \bibinfo {author} {\bibfnamefont {V.}~\bibnamefont {Bisogni}}, \bibinfo {author} {\bibfnamefont {W.}~\bibnamefont {He}}, \bibinfo {author} {\bibfnamefont {X.}~\bibnamefont {Luo}}, \bibinfo {author} {\bibfnamefont {M.~P.~M.}\ \bibnamefont {Dean}}, \bibinfo {author} {\bibfnamefont {M.~K.}\ \bibnamefont {Lajer}}, \bibinfo {author} {\bibfnamefont {M.}~\bibnamefont {Kagan}}, \bibinfo {author} {\bibfnamefont {J.~J.}\ \bibnamefont {Turner}}, \bibinfo {author} {\bibfnamefont {Y.}~\bibnamefont {Cheng}},\ and\ \bibinfo {author} {\bibfnamefont {S.}~\bibnamefont {Gasiorowski}},\ }\bibfield  {title} {\bibinfo {title} {Data and analysis materials for posterior inference of hamiltonian parameters from rixs spectroscopy},\ }\href {https://doi.org/10.5281/zenodo.21895817} {10.5281/zenodo.21895817} (\bibinfo {year} {2026}{\natexlab{a}})\BibitemShut {NoStop}%
\bibitem [{\citenamefont {Zaanen}\ \emph {et~al.}(1985)\citenamefont {Zaanen}, \citenamefont {Sawatzky},\ and\ \citenamefont {Allen}}]{Zaanen1985}%
  \BibitemOpen
  \bibfield  {author} {\bibinfo {author} {\bibfnamefont {J.}~\bibnamefont {Zaanen}}, \bibinfo {author} {\bibfnamefont {G.~A.}\ \bibnamefont {Sawatzky}},\ and\ \bibinfo {author} {\bibfnamefont {J.~W.}\ \bibnamefont {Allen}},\ }\bibfield  {title} {\bibinfo {title} {Band gaps and electronic structure of transition-metal compounds},\ }\href {https://doi.org/10.1103/PhysRevLett.55.418} {\bibfield  {journal} {\bibinfo  {journal} {Physical Review Letters}\ }\textbf {\bibinfo {volume} {55}},\ \bibinfo {pages} {418} (\bibinfo {year} {1985})}\BibitemShut {NoStop}%
\bibitem [{\citenamefont {Klein}\ \emph {et~al.}(2026{\natexlab{b}})\citenamefont {Klein}, \citenamefont {Neiswanger}, \citenamefont {Ratner}, \citenamefont {Kagan},\ and\ \citenamefont {Gasiorowski}}]{sbisupercharged}%
  \BibitemOpen
  \bibfield  {author} {\bibinfo {author} {\bibfnamefont {S.}~\bibnamefont {Klein}}, \bibinfo {author} {\bibfnamefont {W.}~\bibnamefont {Neiswanger}}, \bibinfo {author} {\bibfnamefont {D.}~\bibnamefont {Ratner}}, \bibinfo {author} {\bibfnamefont {M.}~\bibnamefont {Kagan}},\ and\ \bibinfo {author} {\bibfnamefont {S.}~\bibnamefont {Gasiorowski}},\ }\href {https://arxiv.org/abs/2602.06900} {\bibinfo {title} {Supercharging simulation-based inference for bayesian optimal experimental design}} (\bibinfo {year} {2026}{\natexlab{b}}),\ \Eprint {https://arxiv.org/abs/2602.06900} {arXiv:2602.06900 [cs.LG]} \BibitemShut {NoStop}%
\bibitem [{\citenamefont {M{\"o}lder}\ \emph {et~al.}(2021)\citenamefont {M{\"o}lder}, \citenamefont {Jablonski}, \citenamefont {Letcher}, \citenamefont {Hall}, \citenamefont {Tomkins-Tinch}, \citenamefont {Sochat}, \citenamefont {Forster}, \citenamefont {Lee}, \citenamefont {Twardziok}, \citenamefont {Kanitz}, \citenamefont {Wilm}, \citenamefont {Holtgrewe}, \citenamefont {Rahmann}, \citenamefont {Nahnsen},\ and\ \citenamefont {K{\"o}ster}}]{snakemake}%
  \BibitemOpen
  \bibfield  {author} {\bibinfo {author} {\bibfnamefont {F.}~\bibnamefont {M{\"o}lder}}, \bibinfo {author} {\bibfnamefont {K.~P.}\ \bibnamefont {Jablonski}}, \bibinfo {author} {\bibfnamefont {B.}~\bibnamefont {Letcher}}, \bibinfo {author} {\bibfnamefont {M.~B.}\ \bibnamefont {Hall}}, \bibinfo {author} {\bibfnamefont {C.~H.}\ \bibnamefont {Tomkins-Tinch}}, \bibinfo {author} {\bibfnamefont {V.}~\bibnamefont {Sochat}}, \bibinfo {author} {\bibfnamefont {J.}~\bibnamefont {Forster}}, \bibinfo {author} {\bibfnamefont {S.}~\bibnamefont {Lee}}, \bibinfo {author} {\bibfnamefont {S.~O.}\ \bibnamefont {Twardziok}}, \bibinfo {author} {\bibfnamefont {A.}~\bibnamefont {Kanitz}}, \bibinfo {author} {\bibfnamefont {A.}~\bibnamefont {Wilm}}, \bibinfo {author} {\bibfnamefont {M.}~\bibnamefont {Holtgrewe}}, \bibinfo {author} {\bibfnamefont {S.}~\bibnamefont {Rahmann}}, \bibinfo {author} {\bibfnamefont {S.}~\bibnamefont {Nahnsen}},\ and\ \bibinfo {author} {\bibfnamefont {J.}~\bibnamefont {K{\"o}ster}},\ }\bibfield  {title} {\bibinfo {title} {Sustainable data analysis with {Snakemake}},\ }\href {https://doi.org/10.12688/f1000research.29032.2} {\bibfield  {journal} {\bibinfo  {journal} {F1000Research}\ }\textbf {\bibinfo {volume} {10}},\ \bibinfo {pages} {33} (\bibinfo {year} {2021})}\BibitemShut {NoStop}%
\bibitem [{\citenamefont {Arts}\ \emph {et~al.}(2023)\citenamefont {Arts}, \citenamefont {Zalmstra}, \citenamefont {Vollprecht}, \citenamefont {de~Jager}, \citenamefont {Morcotilo},\ and\ \citenamefont {Hofer}}]{pixi}%
  \BibitemOpen
  \bibfield  {author} {\bibinfo {author} {\bibfnamefont {R.}~\bibnamefont {Arts}}, \bibinfo {author} {\bibfnamefont {B.}~\bibnamefont {Zalmstra}}, \bibinfo {author} {\bibfnamefont {W.}~\bibnamefont {Vollprecht}}, \bibinfo {author} {\bibfnamefont {T.}~\bibnamefont {de~Jager}}, \bibinfo {author} {\bibfnamefont {N.}~\bibnamefont {Morcotilo}},\ and\ \bibinfo {author} {\bibfnamefont {J.}~\bibnamefont {Hofer}},\ }\href {https://pixi.sh} {\bibinfo {title} {pixi: A cross-platform, language agnostic, package/project management tool for development in virtual environments}},\ \bibinfo {howpublished} {GitHub} (\bibinfo {year} {2023})\BibitemShut {NoStop}%
\bibitem [{\citenamefont {Loshchilov}\ and\ \citenamefont {Hutter}(2019)}]{Loshchilov2019}%
  \BibitemOpen
  \bibfield  {author} {\bibinfo {author} {\bibfnamefont {I.}~\bibnamefont {Loshchilov}}\ and\ \bibinfo {author} {\bibfnamefont {F.}~\bibnamefont {Hutter}},\ }\bibfield  {title} {\bibinfo {title} {Decoupled weight decay regularization},\ }in\ \href@noop {} {\emph {\bibinfo {booktitle} {International Conference on Learning Representations (ICLR)}}}\ (\bibinfo {year} {2019})\BibitemShut {NoStop}%
\bibitem [{\citenamefont {Talts}\ \emph {et~al.}(2018)\citenamefont {Talts}, \citenamefont {Betancourt}, \citenamefont {Simpson}, \citenamefont {Vehtari},\ and\ \citenamefont {Gelman}}]{Talts2018}%
  \BibitemOpen
  \bibfield  {author} {\bibinfo {author} {\bibfnamefont {S.}~\bibnamefont {Talts}}, \bibinfo {author} {\bibfnamefont {M.}~\bibnamefont {Betancourt}}, \bibinfo {author} {\bibfnamefont {D.}~\bibnamefont {Simpson}}, \bibinfo {author} {\bibfnamefont {A.}~\bibnamefont {Vehtari}},\ and\ \bibinfo {author} {\bibfnamefont {A.}~\bibnamefont {Gelman}},\ }\href@noop {} {\bibinfo {title} {{Validating Bayesian Inference Algorithms with Simulation-Based Calibration}}} (\bibinfo {year} {2018}),\ \Eprint {https://arxiv.org/abs/1804.06788} {arXiv:1804.06788 [stat.ME]} \BibitemShut {NoStop}%
\bibitem [{\citenamefont {Van~Rossum}\ and\ \citenamefont {Drake}(2009)}]{python3}%
  \BibitemOpen
  \bibfield  {author} {\bibinfo {author} {\bibfnamefont {G.}~\bibnamefont {Van~Rossum}}\ and\ \bibinfo {author} {\bibfnamefont {F.~L.}\ \bibnamefont {Drake}},\ }\bibfield  {title} {\bibinfo {title} {Python 3 reference manual},\ }\href@noop {} {\  (\bibinfo {year} {2009})}\BibitemShut {NoStop}%
\bibitem [{\citenamefont {Paszke}\ \emph {et~al.}(2019)\citenamefont {Paszke}, \citenamefont {Gross}, \citenamefont {Massa}, \citenamefont {Lerer}, \citenamefont {Bradbury}, \citenamefont {Chanan}, \citenamefont {Killeen}, \citenamefont {Lin}, \citenamefont {Gimelshein}, \citenamefont {Antiga}, \citenamefont {Desmaison}, \citenamefont {K{\"o}pf}, \citenamefont {Yang}, \citenamefont {DeVito}, \citenamefont {Raison}, \citenamefont {Tejani}, \citenamefont {Chilamkurthy}, \citenamefont {Steiner}, \citenamefont {Fang}, \citenamefont {Bai},\ and\ \citenamefont {Chintala}}]{pytorch}%
  \BibitemOpen
  \bibfield  {author} {\bibinfo {author} {\bibfnamefont {A.}~\bibnamefont {Paszke}}, \bibinfo {author} {\bibfnamefont {S.}~\bibnamefont {Gross}}, \bibinfo {author} {\bibfnamefont {F.}~\bibnamefont {Massa}}, \bibinfo {author} {\bibfnamefont {A.}~\bibnamefont {Lerer}}, \bibinfo {author} {\bibfnamefont {J.}~\bibnamefont {Bradbury}}, \bibinfo {author} {\bibfnamefont {G.}~\bibnamefont {Chanan}}, \bibinfo {author} {\bibfnamefont {T.}~\bibnamefont {Killeen}}, \bibinfo {author} {\bibfnamefont {Z.}~\bibnamefont {Lin}}, \bibinfo {author} {\bibfnamefont {N.}~\bibnamefont {Gimelshein}}, \bibinfo {author} {\bibfnamefont {L.}~\bibnamefont {Antiga}}, \bibinfo {author} {\bibfnamefont {A.}~\bibnamefont {Desmaison}}, \bibinfo {author} {\bibfnamefont {A.}~\bibnamefont {K{\"o}pf}}, \bibinfo {author} {\bibfnamefont {E.}~\bibnamefont {Yang}}, \bibinfo {author} {\bibfnamefont {Z.}~\bibnamefont {DeVito}}, \bibinfo {author} {\bibfnamefont {M.}~\bibnamefont {Raison}}, \bibinfo {author} {\bibfnamefont {A.}~\bibnamefont {Tejani}}, \bibinfo {author} {\bibfnamefont {S.}~\bibnamefont {Chilamkurthy}}, \bibinfo {author} {\bibfnamefont {B.}~\bibnamefont {Steiner}}, \bibinfo {author} {\bibfnamefont {L.}~\bibnamefont {Fang}}, \bibinfo {author} {\bibfnamefont {J.}~\bibnamefont {Bai}},\ and\ \bibinfo {author} {\bibfnamefont {S.}~\bibnamefont {Chintala}},\ }\bibfield  {title} {\bibinfo {title} {{PyTorch}: An imperative style, high-performance deep learning library},\ }in\ \href {https://papers.nips.cc/paper/9015-pytorch-an-imperative-style-high-performance-deep-learning-library} {\emph {\bibinfo {booktitle} {Advances in Neural Information Processing Systems}}},\ Vol.~\bibinfo {volume} {32}\ (\bibinfo {year} {2019})\ pp.\ \bibinfo {pages} {8024--8035}\BibitemShut {NoStop}%
\bibitem [{\citenamefont {Tejero-Cantero}\ \emph {et~al.}(2020)\citenamefont {Tejero-Cantero}, \citenamefont {Boelts}, \citenamefont {Deistler}, \citenamefont {Lueckmann}, \citenamefont {Durkan}, \citenamefont {Gon{\c{c}}alves}, \citenamefont {Greenberg},\ and\ \citenamefont {Macke}}]{sbi_toolkit}%
  \BibitemOpen
  \bibfield  {author} {\bibinfo {author} {\bibfnamefont {A.}~\bibnamefont {Tejero-Cantero}}, \bibinfo {author} {\bibfnamefont {J.}~\bibnamefont {Boelts}}, \bibinfo {author} {\bibfnamefont {M.}~\bibnamefont {Deistler}}, \bibinfo {author} {\bibfnamefont {J.-M.}\ \bibnamefont {Lueckmann}}, \bibinfo {author} {\bibfnamefont {C.}~\bibnamefont {Durkan}}, \bibinfo {author} {\bibfnamefont {P.~J.}\ \bibnamefont {Gon{\c{c}}alves}}, \bibinfo {author} {\bibfnamefont {D.~S.}\ \bibnamefont {Greenberg}},\ and\ \bibinfo {author} {\bibfnamefont {J.~H.}\ \bibnamefont {Macke}},\ }\bibfield  {title} {\bibinfo {title} {sbi: A toolkit for simulation-based inference},\ }\href {https://doi.org/10.21105/joss.02505} {\bibfield  {journal} {\bibinfo  {journal} {Journal of Open Source Software}\ }\textbf {\bibinfo {volume} {5}},\ \bibinfo {pages} {2505} (\bibinfo {year} {2020})}\BibitemShut {NoStop}%
\bibitem [{\citenamefont {Boelts}\ \emph {et~al.}(2025)\citenamefont {Boelts}, \citenamefont {Deistler}, \citenamefont {Gloeckler}, \citenamefont {Tejero-Cantero}, \citenamefont {Lueckmann}, \citenamefont {Moss}, \citenamefont {Steinbach}, \citenamefont {Moreau}, \citenamefont {Muratore}, \citenamefont {Linhart}, \citenamefont {Durkan}, \citenamefont {Vetter}, \citenamefont {Miller}, \citenamefont {Herold}, \citenamefont {Ziaeemehr}, \citenamefont {Pals}, \citenamefont {Gruner}, \citenamefont {Bischoff}, \citenamefont {Krouglova}, \citenamefont {Gao}, \citenamefont {Lappalainen}, \citenamefont {Mucs{\'a}nyi}, \citenamefont {Pei}, \citenamefont {Schulz}, \citenamefont {Stefanidi}, \citenamefont {Rodrigues}, \citenamefont {Schr{\"o}der}, \citenamefont {Zaid}, \citenamefont {Beck}, \citenamefont {Kapoor}, \citenamefont {Greenberg}, \citenamefont {Gon{\c{c}}alves},\ and\ \citenamefont {Macke}}]{sbi_reloaded}%
  \BibitemOpen
  \bibfield  {author} {\bibinfo {author} {\bibfnamefont {J.}~\bibnamefont {Boelts}}, \bibinfo {author} {\bibfnamefont {M.}~\bibnamefont {Deistler}}, \bibinfo {author} {\bibfnamefont {M.}~\bibnamefont {Gloeckler}}, \bibinfo {author} {\bibfnamefont {{\'A}.}~\bibnamefont {Tejero-Cantero}}, \bibinfo {author} {\bibfnamefont {J.-M.}\ \bibnamefont {Lueckmann}}, \bibinfo {author} {\bibfnamefont {G.}~\bibnamefont {Moss}}, \bibinfo {author} {\bibfnamefont {P.}~\bibnamefont {Steinbach}}, \bibinfo {author} {\bibfnamefont {T.}~\bibnamefont {Moreau}}, \bibinfo {author} {\bibfnamefont {F.}~\bibnamefont {Muratore}}, \bibinfo {author} {\bibfnamefont {J.}~\bibnamefont {Linhart}}, \bibinfo {author} {\bibfnamefont {C.}~\bibnamefont {Durkan}}, \bibinfo {author} {\bibfnamefont {J.}~\bibnamefont {Vetter}}, \bibinfo {author} {\bibfnamefont {B.~K.}\ \bibnamefont {Miller}}, \bibinfo {author} {\bibfnamefont {M.}~\bibnamefont {Herold}}, \bibinfo {author} {\bibfnamefont {A.}~\bibnamefont {Ziaeemehr}}, \bibinfo {author} {\bibfnamefont {M.}~\bibnamefont {Pals}}, \bibinfo {author} {\bibfnamefont {T.}~\bibnamefont {Gruner}}, \bibinfo {author} {\bibfnamefont {S.}~\bibnamefont {Bischoff}}, \bibinfo {author} {\bibfnamefont {N.}~\bibnamefont {Krouglova}}, \bibinfo {author} {\bibfnamefont {R.}~\bibnamefont {Gao}}, \bibinfo {author} {\bibfnamefont {J.~K.}\ \bibnamefont {Lappalainen}}, \bibinfo {author} {\bibfnamefont {B.}~\bibnamefont {Mucs{\'a}nyi}}, \bibinfo {author} {\bibfnamefont {F.}~\bibnamefont {Pei}}, \bibinfo {author} {\bibfnamefont {A.}~\bibnamefont {Schulz}}, \bibinfo {author} {\bibfnamefont {Z.}~\bibnamefont {Stefanidi}}, \bibinfo {author} {\bibfnamefont {P.}~\bibnamefont {Rodrigues}}, \bibinfo {author} {\bibfnamefont {C.}~\bibnamefont {Schr{\"o}der}}, \bibinfo {author} {\bibfnamefont {F.~A.}\ \bibnamefont {Zaid}}, \bibinfo {author} {\bibfnamefont {J.}~\bibnamefont {Beck}}, \bibinfo {author} {\bibfnamefont {J.}~\bibnamefont {Kapoor}}, \bibinfo {author} {\bibfnamefont {D.~S.}\ \bibnamefont {Greenberg}}, \bibinfo {author} {\bibfnamefont {P.~J.}\ \bibnamefont {Gon{\c{c}}alves}},\ and\ \bibinfo {author} {\bibfnamefont {J.~H.}\ \bibnamefont {Macke}},\ }\bibfield  {title} {\bibinfo {title} {sbi reloaded: a toolkit for simulation-based inference workflows},\ }\href {https://doi.org/10.21105/joss.07754} {\bibfield  {journal} {\bibinfo  {journal} {Journal of Open Source Software}\ }\textbf {\bibinfo {volume} {10}},\ \bibinfo {pages} {7754} (\bibinfo {year} {2025})}\BibitemShut {NoStop}%
\bibitem [{\citenamefont {Durkan}\ \emph {et~al.}(2020)\citenamefont {Durkan}, \citenamefont {Bekasov}, \citenamefont {Murray},\ and\ \citenamefont {Papamakarios}}]{nflows}%
  \BibitemOpen
  \bibfield  {author} {\bibinfo {author} {\bibfnamefont {C.}~\bibnamefont {Durkan}}, \bibinfo {author} {\bibfnamefont {A.}~\bibnamefont {Bekasov}}, \bibinfo {author} {\bibfnamefont {I.}~\bibnamefont {Murray}},\ and\ \bibinfo {author} {\bibfnamefont {G.}~\bibnamefont {Papamakarios}},\ }\href {https://doi.org/10.5281/zenodo.4296287} {\bibinfo {title} {{nflows}: normalizing flows in {PyTorch}}} (\bibinfo {year} {2020})\BibitemShut {NoStop}%
\bibitem [{\citenamefont {Harris}\ \emph {et~al.}(2020)\citenamefont {Harris}, \citenamefont {Millman}, \citenamefont {van~der Walt}, \citenamefont {Gommers}, \citenamefont {Virtanen}, \citenamefont {Cournapeau}, \citenamefont {Wieser}, \citenamefont {Taylor}, \citenamefont {Berg}, \citenamefont {Smith}, \citenamefont {Kern}, \citenamefont {Picus}, \citenamefont {Hoyer}, \citenamefont {van Kerkwijk}, \citenamefont {Brett}, \citenamefont {Haldane}, \citenamefont {del R{\'i}o}, \citenamefont {Wiebe}, \citenamefont {Peterson}, \citenamefont {G{\'e}rard-Marchant}, \citenamefont {Sheppard}, \citenamefont {Reddy}, \citenamefont {Weckesser}, \citenamefont {Abbasi}, \citenamefont {Gohlke},\ and\ \citenamefont {Oliphant}}]{numpy}%
  \BibitemOpen
  \bibfield  {author} {\bibinfo {author} {\bibfnamefont {C.~R.}\ \bibnamefont {Harris}}, \bibinfo {author} {\bibfnamefont {K.~J.}\ \bibnamefont {Millman}}, \bibinfo {author} {\bibfnamefont {S.~J.}\ \bibnamefont {van~der Walt}}, \bibinfo {author} {\bibfnamefont {R.}~\bibnamefont {Gommers}}, \bibinfo {author} {\bibfnamefont {P.}~\bibnamefont {Virtanen}}, \bibinfo {author} {\bibfnamefont {D.}~\bibnamefont {Cournapeau}}, \bibinfo {author} {\bibfnamefont {E.}~\bibnamefont {Wieser}}, \bibinfo {author} {\bibfnamefont {J.}~\bibnamefont {Taylor}}, \bibinfo {author} {\bibfnamefont {S.}~\bibnamefont {Berg}}, \bibinfo {author} {\bibfnamefont {N.~J.}\ \bibnamefont {Smith}}, \bibinfo {author} {\bibfnamefont {R.}~\bibnamefont {Kern}}, \bibinfo {author} {\bibfnamefont {M.}~\bibnamefont {Picus}}, \bibinfo {author} {\bibfnamefont {S.}~\bibnamefont {Hoyer}}, \bibinfo {author} {\bibfnamefont {M.~H.}\ \bibnamefont {van Kerkwijk}}, \bibinfo {author} {\bibfnamefont {M.}~\bibnamefont {Brett}}, \bibinfo {author} {\bibfnamefont {A.}~\bibnamefont {Haldane}}, \bibinfo {author} {\bibfnamefont {J.~F.}\ \bibnamefont {del R{\'i}o}}, \bibinfo {author} {\bibfnamefont {M.}~\bibnamefont {Wiebe}}, \bibinfo {author} {\bibfnamefont {P.}~\bibnamefont {Peterson}}, \bibinfo {author} {\bibfnamefont {P.}~\bibnamefont {G{\'e}rard-Marchant}}, \bibinfo {author} {\bibfnamefont {K.}~\bibnamefont {Sheppard}}, \bibinfo {author} {\bibfnamefont {T.}~\bibnamefont {Reddy}}, \bibinfo {author} {\bibfnamefont {W.}~\bibnamefont {Weckesser}}, \bibinfo {author} {\bibfnamefont {H.}~\bibnamefont {Abbasi}}, \bibinfo {author} {\bibfnamefont {C.}~\bibnamefont {Gohlke}},\ and\ \bibinfo {author} {\bibfnamefont {T.~E.}\ \bibnamefont {Oliphant}},\ }\bibfield  {title} {\bibinfo {title} {Array programming with {NumPy}},\ }\href {https://doi.org/10.1038/s41586-020-2649-2} {\bibfield  {journal} {\bibinfo  {journal} {Nature}\ }\textbf {\bibinfo {volume} {585}},\ \bibinfo {pages} {357} (\bibinfo {year} {2020})}\BibitemShut {NoStop}%
\bibitem [{\citenamefont {Virtanen}\ \emph {et~al.}(2020)\citenamefont {Virtanen}, \citenamefont {Gommers}, \citenamefont {Oliphant}, \citenamefont {Haberland}, \citenamefont {Reddy}, \citenamefont {Cournapeau}, \citenamefont {Burovski}, \citenamefont {Peterson}, \citenamefont {Weckesser}, \citenamefont {Bright}, \citenamefont {van~der Walt}, \citenamefont {Brett}, \citenamefont {Wilson}, \citenamefont {Millman}, \citenamefont {Mayorov}, \citenamefont {Nelson}, \citenamefont {Jones}, \citenamefont {Kern}, \citenamefont {Larson}, \citenamefont {Carey}, \citenamefont {Polat}, \citenamefont {Feng}, \citenamefont {Moore}, \citenamefont {VanderPlas}, \citenamefont {Laxalde}, \citenamefont {Perktold}, \citenamefont {Cimrman}, \citenamefont {Henriksen}, \citenamefont {Quintero}, \citenamefont {Harris}, \citenamefont {Archibald}, \citenamefont {Ribeiro}, \citenamefont {Pedregosa}, \citenamefont {van Mulbregt},\ and\ \citenamefont {{SciPy 1.0 Contributors}}}]{scipy}%
  \BibitemOpen
  \bibfield  {author} {\bibinfo {author} {\bibfnamefont {P.}~\bibnamefont {Virtanen}}, \bibinfo {author} {\bibfnamefont {R.}~\bibnamefont {Gommers}}, \bibinfo {author} {\bibfnamefont {T.~E.}\ \bibnamefont {Oliphant}}, \bibinfo {author} {\bibfnamefont {M.}~\bibnamefont {Haberland}}, \bibinfo {author} {\bibfnamefont {T.}~\bibnamefont {Reddy}}, \bibinfo {author} {\bibfnamefont {D.}~\bibnamefont {Cournapeau}}, \bibinfo {author} {\bibfnamefont {E.}~\bibnamefont {Burovski}}, \bibinfo {author} {\bibfnamefont {P.}~\bibnamefont {Peterson}}, \bibinfo {author} {\bibfnamefont {W.}~\bibnamefont {Weckesser}}, \bibinfo {author} {\bibfnamefont {J.}~\bibnamefont {Bright}}, \bibinfo {author} {\bibfnamefont {S.~J.}\ \bibnamefont {van~der Walt}}, \bibinfo {author} {\bibfnamefont {M.}~\bibnamefont {Brett}}, \bibinfo {author} {\bibfnamefont {J.}~\bibnamefont {Wilson}}, \bibinfo {author} {\bibfnamefont {K.~J.}\ \bibnamefont {Millman}}, \bibinfo {author} {\bibfnamefont {N.}~\bibnamefont {Mayorov}}, \bibinfo {author} {\bibfnamefont {A.~R.~J.}\ \bibnamefont {Nelson}}, \bibinfo {author} {\bibfnamefont {E.}~\bibnamefont {Jones}}, \bibinfo {author} {\bibfnamefont {R.}~\bibnamefont {Kern}}, \bibinfo {author} {\bibfnamefont {E.}~\bibnamefont {Larson}}, \bibinfo {author} {\bibfnamefont {C.~J.}\ \bibnamefont {Carey}}, \bibinfo {author} {\bibfnamefont {{\.I}.}~\bibnamefont {Polat}}, \bibinfo {author} {\bibfnamefont {Y.}~\bibnamefont {Feng}}, \bibinfo {author} {\bibfnamefont {E.~W.}\ \bibnamefont {Moore}}, \bibinfo {author} {\bibfnamefont {J.}~\bibnamefont {VanderPlas}}, \bibinfo {author} {\bibfnamefont {D.}~\bibnamefont {Laxalde}}, \bibinfo {author} {\bibfnamefont {J.}~\bibnamefont {Perktold}}, \bibinfo {author} {\bibfnamefont {R.}~\bibnamefont {Cimrman}}, \bibinfo {author} {\bibfnamefont {I.}~\bibnamefont {Henriksen}}, \bibinfo {author} {\bibfnamefont {E.~A.}\ \bibnamefont {Quintero}}, \bibinfo {author} {\bibfnamefont {C.~R.}\ \bibnamefont {Harris}}, \bibinfo {author} {\bibfnamefont {A.~M.}\ \bibnamefont {Archibald}}, \bibinfo {author} {\bibfnamefont {A.~H.}\ \bibnamefont {Ribeiro}}, \bibinfo {author} {\bibfnamefont {F.}~\bibnamefont {Pedregosa}}, \bibinfo {author} {\bibfnamefont {P.}~\bibnamefont {van Mulbregt}},\ and\ \bibinfo {author} {\bibnamefont {{SciPy 1.0 Contributors}}},\ }\bibfield  {title} {\bibinfo {title} {{SciPy} 1.0: Fundamental algorithms for scientific computing in {Python}},\ }\href {https://doi.org/10.1038/s41592-019-0686-2} {\bibfield  {journal} {\bibinfo  {journal} {Nature Methods}\ }\textbf {\bibinfo {volume} {17}},\ \bibinfo {pages} {261} (\bibinfo {year} {2020})}\BibitemShut {NoStop}%
\bibitem [{\citenamefont {Pedregosa}\ \emph {et~al.}(2011)\citenamefont {Pedregosa}, \citenamefont {Varoquaux}, \citenamefont {Gramfort}, \citenamefont {Michel}, \citenamefont {Thirion}, \citenamefont {Grisel}, \citenamefont {Blondel}, \citenamefont {Prettenhofer}, \citenamefont {Weiss}, \citenamefont {Dubourg}, \citenamefont {Vanderplas}, \citenamefont {Passos}, \citenamefont {Cournapeau}, \citenamefont {Brucher}, \citenamefont {Perrot},\ and\ \citenamefont {Duchesnay}}]{scikitlearn}%
  \BibitemOpen
  \bibfield  {author} {\bibinfo {author} {\bibfnamefont {F.}~\bibnamefont {Pedregosa}}, \bibinfo {author} {\bibfnamefont {G.}~\bibnamefont {Varoquaux}}, \bibinfo {author} {\bibfnamefont {A.}~\bibnamefont {Gramfort}}, \bibinfo {author} {\bibfnamefont {V.}~\bibnamefont {Michel}}, \bibinfo {author} {\bibfnamefont {B.}~\bibnamefont {Thirion}}, \bibinfo {author} {\bibfnamefont {O.}~\bibnamefont {Grisel}}, \bibinfo {author} {\bibfnamefont {M.}~\bibnamefont {Blondel}}, \bibinfo {author} {\bibfnamefont {P.}~\bibnamefont {Prettenhofer}}, \bibinfo {author} {\bibfnamefont {R.}~\bibnamefont {Weiss}}, \bibinfo {author} {\bibfnamefont {V.}~\bibnamefont {Dubourg}}, \bibinfo {author} {\bibfnamefont {J.}~\bibnamefont {Vanderplas}}, \bibinfo {author} {\bibfnamefont {A.}~\bibnamefont {Passos}}, \bibinfo {author} {\bibfnamefont {D.}~\bibnamefont {Cournapeau}}, \bibinfo {author} {\bibfnamefont {M.}~\bibnamefont {Brucher}}, \bibinfo {author} {\bibfnamefont {M.}~\bibnamefont {Perrot}},\ and\ \bibinfo {author} {\bibfnamefont {E.}~\bibnamefont {Duchesnay}},\ }\bibfield  {title} {\bibinfo {title} {Scikit-learn: Machine learning in {Python}},\ }\href {https://jmlr.org/papers/v12/pedregosa11a.html} {\bibfield  {journal} {\bibinfo  {journal} {Journal of Machine Learning Research}\ }\textbf {\bibinfo {volume} {12}},\ \bibinfo {pages} {2825} (\bibinfo {year} {2011})}\BibitemShut {NoStop}%
\bibitem [{\citenamefont {Newville}\ \emph {et~al.}(2014)\citenamefont {Newville}, \citenamefont {Stensitzki}, \citenamefont {Allen},\ and\ \citenamefont {Ingargiola}}]{lmfit}%
  \BibitemOpen
  \bibfield  {author} {\bibinfo {author} {\bibfnamefont {M.}~\bibnamefont {Newville}}, \bibinfo {author} {\bibfnamefont {T.}~\bibnamefont {Stensitzki}}, \bibinfo {author} {\bibfnamefont {D.~B.}\ \bibnamefont {Allen}},\ and\ \bibinfo {author} {\bibfnamefont {A.}~\bibnamefont {Ingargiola}},\ }\href {https://doi.org/10.5281/zenodo.11813} {\bibinfo {title} {{LMFIT}: Non-linear least-square minimization and curve-fitting for {Python}}} (\bibinfo {year} {2014})\BibitemShut {NoStop}%
\bibitem [{\citenamefont {Collette}(2013)}]{h5py}%
  \BibitemOpen
  \bibfield  {author} {\bibinfo {author} {\bibfnamefont {A.}~\bibnamefont {Collette}},\ }\href@noop {} {\emph {\bibinfo {title} {Python and {HDF5}}}}\ (\bibinfo  {publisher} {O'Reilly Media},\ \bibinfo {address} {Sebastopol, CA},\ \bibinfo {year} {2013})\BibitemShut {NoStop}%
\bibitem [{\citenamefont {Hoyer}\ and\ \citenamefont {Hamman}(2017)}]{xarray}%
  \BibitemOpen
  \bibfield  {author} {\bibinfo {author} {\bibfnamefont {S.}~\bibnamefont {Hoyer}}\ and\ \bibinfo {author} {\bibfnamefont {J.~J.}\ \bibnamefont {Hamman}},\ }\bibfield  {title} {\bibinfo {title} {xarray: {N-D} labeled arrays and datasets in {Python}},\ }\href {https://doi.org/10.5334/jors.148} {\bibfield  {journal} {\bibinfo  {journal} {Journal of Open Research Software}\ }\textbf {\bibinfo {volume} {5}},\ \bibinfo {pages} {10} (\bibinfo {year} {2017})}\BibitemShut {NoStop}%
\bibitem [{\citenamefont {McKinney}(2010)}]{pandas}%
  \BibitemOpen
  \bibfield  {author} {\bibinfo {author} {\bibfnamefont {W.}~\bibnamefont {McKinney}},\ }\bibfield  {title} {\bibinfo {title} {Data structures for statistical computing in {Python}},\ }in\ \href {https://doi.org/10.25080/Majora-92bf1922-00a} {\emph {\bibinfo {booktitle} {Proceedings of the 9th Python in Science Conference}}},\ \bibinfo {editor} {edited by\ \bibinfo {editor} {\bibfnamefont {S.}~\bibnamefont {van~der Walt}}\ and\ \bibinfo {editor} {\bibfnamefont {J.}~\bibnamefont {Millman}}}\ (\bibinfo {year} {2010})\ pp.\ \bibinfo {pages} {56--61}\BibitemShut {NoStop}%
\bibitem [{\citenamefont {Hunter}(2007)}]{matplotlib}%
  \BibitemOpen
  \bibfield  {author} {\bibinfo {author} {\bibfnamefont {J.~D.}\ \bibnamefont {Hunter}},\ }\bibfield  {title} {\bibinfo {title} {Matplotlib: A 2{D} graphics environment},\ }\href {https://doi.org/10.1109/MCSE.2007.55} {\bibfield  {journal} {\bibinfo  {journal} {Computing in Science \& Engineering}\ }\textbf {\bibinfo {volume} {9}},\ \bibinfo {pages} {90} (\bibinfo {year} {2007})}\BibitemShut {NoStop}%
\bibitem [{\citenamefont {Yadan}(2019)}]{hydra}%
  \BibitemOpen
  \bibfield  {author} {\bibinfo {author} {\bibfnamefont {O.}~\bibnamefont {Yadan}},\ }\href {https://github.com/facebookresearch/hydra} {\bibinfo {title} {Hydra --- a framework for elegantly configuring complex applications}},\ \bibinfo {howpublished} {GitHub} (\bibinfo {year} {2019})\BibitemShut {NoStop}%
\bibitem [{\citenamefont {Biewald}(2020)}]{wandb}%
  \BibitemOpen
  \bibfield  {author} {\bibinfo {author} {\bibfnamefont {L.}~\bibnamefont {Biewald}},\ }\href {https://www.wandb.com/} {\bibinfo {title} {Experiment tracking with {Weights and Biases}}},\ \bibinfo {howpublished} {Software available from wandb.com} (\bibinfo {year} {2020})\BibitemShut {NoStop}%
\end{thebibliography}%

\end{document}